\documentclass{emulateapj}
\usepackage{graphicx, amsmath, amssymb, amsthm}

\makeatother

\newcommand\Ms{M$_\odot$}

\begin{document}

\title{Radiative and Kinetic Feedback by Low-Mass Primordial Stars}
\author{Daniel Whalen\altaffilmark{1,2}, Robert M. Hueckstaedt \altaffilmark{3} \&
Thomas O. McConkie \altaffilmark{2,4}}
\altaffiltext{1}{McWilliams Fellow, Department of Physics, Carnegie Mellon University,
Pittsburgh, PA 15213. Email: dwhalen@lanl.gov}
\altaffiltext{2}{Nuclear \& Particle Physics, Astrophysics, and Cosmology (T-2), Los 
Alamos National Laboratory, Los Alamos, NM 87545}
\altaffiltext{3}{Applied Physics (X-2), Los Alamos National Laboratory}
\altaffiltext{4}{Department of Physics and Astronomy, Brigham Young University, Provo,
UT 84602}

\begin{abstract} 

Ionizing UV radiation and supernova flows amidst clustered minihalos 
at high redshift regulated the rise of the first stellar populations
in the universe.  Previous studies have addressed the effects of 
very massive primordial stars on the collapse of nearby halos into 
new stars, but the absence of the odd-even nucleosynthetic signature 
of pair-instability supernovae in ancient metal-poor stars suggests 
that Population III stars may have been less than 100 M$_{\odot}$.  
We extend our earlier survey of local UV feedback on star formation 
to 25 - 80 M$_{\odot}$ stars and include kinetic feedback by supernovae 
for 25 - 40 M$_{\odot}$ stars.  We find radiative feedback to be 
relatively uniform over this mass range, primarily because the larger 
fluxes of more massive 
stars are offset by their shorter lifetimes.  Our models demonstrate 
that prior to the rise of global UV backgrounds, Lyman-Werner photons 
from nearby stars cannot prevent halos from forming new stars.  These 
calculations also reveal that violent dynamical instabilities can 
erupt in the UV radiation front enveloping a primordial halo but that 
they ultimately have no effect on the formation of a star.  Finally, 
our simulations suggest that relic H II regions surrounding partially 
evaporated halos may expel Lyman-Werner backgrounds at lower redshifts, 
allowing stars to form that were previously suppressed.  We provide 
fits to radiative and kinetic feedback on star formation for use in
both semianalytic models and numerical simulations.

\end{abstract}


\section{Introduction}

The survival of cosmological minihalos in the ionizing and Lyman-Werner (LW) 
UV fields of primordial stars is key to the rise of stellar populations at high 
redshifts.  Numerical models suggest that the first stars are very massive, 25 -
500 M$_{\odot}$, and that they form in isolation in small dark matter halos of 
$\sim$ 10$^5$ - 10$^7$ \Ms \ at $z \sim$ 20 - 30 \citep{bcl99,bcl02,nu01,abn00,
abn02,on07}.  These stars create large H II regions 2.5 - 5 kpc in radius that 
can engulf nearby halos \citep{wan04,ket04,abs06,awb07,wa08a}.  From z $\sim$ 
10 - 20 Population III (Pop III) stars also build up a global LW background 
that sterilizes minihalos of H$_2$, delaying or preventing the formation of new 
stars 
\citep{hrl97,har00,met01,mbh06,wa07,su07,on08,wa07,jgb07,mbh09,ahn09}. At high 
redshifts, 
ionizing radiation is relatively local while LW photons can originate from 
many megaparsecs away because their energies lie below the ionization limit of H.

The picture is simpler in the first generation, in which there is no LW background.  
Simulations of halo photoevaporation by nearby very massive ($\gtrsim$ 100 \Ms) 
primordial stars have been performed, both with \citep{oet05,su06,as07,wet08b,wa08b,
hus09,suh09} and without \citep{il04,il05} H$_2$ gas chemistry. The two- and 
three-dimensional studies are in good agreement with each other but not with the 
one-dimensional studies \citep{as07}, primarily due to the serious hydrodynamical 
artifacts that arise from unphysical shock focusing and central bounce that occur in
one-dimensional spherical coordinate meshes.  When multifrequency photon transport is 
coupled to primordial chemistry and hydrodynamics, halo photoevaporation proceeds 
in two stages.  First, LW photons from the star reach the halo before the I-front, 
partly dissociating molecular hydrogen in its core and slowing or halting its collapse.  
Not all of the H$_2$ is erradicated because molecular hydrogen deep in the halo partly 
shields itself from the photons and small free electron fractions restore H$_2$ by 
catalysis through the H$^-$ channel. The I-front decelerates as it approaches 
the halo and transforms from R-type to D-type. Molecular hydrogen forms in the small 
ionized fractions and warm temperatures in the outer layers of the front 
\citep[e.g.][]{rgs01,wn08b}, partly shielding the interior of the halo from LW photons 
from the star and even allowing H$_2$ fractions in the core to rise in some cases. The 
ionization front preferentially advances through the low densities above and below the 
halo and assumes a cometary shape, with the inner regions of the halo casting a shadow.

Second, after the death of the star the H II region surrounding the halo begins to 
recombine out of equilibrium, rapidly forming H$_2$ and HD that can cool the ionized 
gas down to the CMB temperature \citep{nu02,jb07,yet07,get08}.  At the same time, H$_2$ 
reconstitutes in the core, rapidly surpassing its original levels and allowing gas to 
again cool and collapse.  Meanwhile, the I-front shock remnant continues to converge 
on the halo core from one side while warm ionized gas presses down into the shadow 
and wraps around the halo from behind.  If the UV flux from the star is not too high, 
the relic shock compresses the core and enriches it with the H$_2$ originally formed 
in the I-front, accelerating its cooling and collapse into a new star.  If the flux 
from the star is large, the ionized gas instead disrupts the core and pre-empts new 
star formation.  The studies performed to date assume stars that do not die in 
supernova (SN) explosions and therefore exclude ram pressure stripping of the halo by 
the remnant \citep{cr08,ss09} and its contamination and cooling by heavy elements. In 
the first generation of stars, both stages of halo photoevaporation occur on time 
scales that are short in comparison to merger or accretion times.  

In large scale calculations of cosmological structure formation, halo evaporation is 
usually modeled with metagalactic ionizing and LW backgrounds that uniformly permeate 
the simulation volume. Such halos evolve very differently than when photon transport 
is performed because they are photoevaporated and photodissociated from the inside 
out.  Rather than being compressed and shielded from LW photons, baryons are expelled 
from all directions.  Consequently, radiative feedback is invariably negative in these 
models.  An exception to this are the recent simulations by \citet{mbh09}, who find 
that low mass halos that are easily evaporated at early times later pool gas and form 
stars when the uniform fields are turned off and the fossil H II regions cool down.  
Self-shielding of H$_2$ from external UV sources cannot be modeled by imposed 
backgrounds, so molecular hydrogen is driven to much lower levels in these halos than 
in real ones.  In general, negative feedback in studies that rely on metagalactic 
UV backgrounds should be taken as a (possibly extreme) upper limit. Star formation in 
the UV environments of the early universe is likely to be significantly more robust.   
 
In principle, the parameter space for local radiative feedback between clustered 
minihalos is vast even if supernova effects are excluded, making the formulation of 
general rules for use in analytical models problematic.  Halo mass, central baryon
density, distance and luminosity of the star, stellar lifetime and spectral profile 
all govern halo evaporation.  Distance and luminosity can be combined in a single 
parameter over intervals in stellar mass for which the shape of the source spectrum 
does not vary much. \citet{wet08b} further reduced this parameter space by examining 
feedback in the smallest halo ever found to form a star in an adaptive mesh refinement 
(AMR) calculation but at four stages of collapse. Since more massive halos at the same 
central baryon density would be less affected by radiation, their findings constitute 
a conservative upper limit to the influence one star can exert on \textit{any} neighbor 
halo capable of forming a star.

In this paper we extend our previous survey by examining the evaporation of minihalos 
by low-mass primordial stars prior to the buildup of a large LW background.  Pop III 
stars from 25 - 80 \Ms \ have spectra with larger LW/ionizing UV ratios than those of 
the 120 \Ms \ star in our previous study and they illuminate other halos for longer 
times.  We sample the same halo and central gas densities as in our earlier study in 
order to place upper limits on the radiative feedback of such stars on more massive 
halos.  By determining the final state of the halo core several Myr after the death 
of the star we construct rules for local radiative and kinetic feedback as a function 
of stellar mass, initial core density, and distance to the star.  In $\S$ 2 we review 
our numerical methods, in $\S$ 3 we describe our cosmological halo models, in $\S$ 4  
we tabulate both radiative and kinetic feedback in a photoevaporated halo, and in $\S$ 
5 we conclude.

\section{Numerical Algorithm}

We perform our halo photoevaporation calculations with ZEUS-MP \citep{het06}, 
a massively-parallel Eulerian astrophysical hydrodynamics code that solves the 
equations of ideal fluid dynamics\footnote{http://lca.ucsd.edu/portal/codes/zeusmp2}:
\vspace{0.1in}
\begin{eqnarray}
\frac{\partial \rho}{\partial t}  & = & - \nabla \: \cdotp \; (\rho {\bf v})  \\
\frac{\partial \rho v_{i}}{\partial t}  & = & - \nabla \: \cdotp \; (\rho v_{i} 
{\bf v}) \: - \: \nabla p \: - \: \rho \nabla \Phi \: - \: \nabla \cdotp {\bf Q}    \\ 
\frac{\partial e}{\partial t}  & = & - \nabla \: \cdotp \; (e {\bf v}) \: - \: p\nabla \: 
\cdotp \: {\bf v} \: - \: \bf{Q} : \nabla  {\bf v}. \\
&  &  \nonumber
\end{eqnarray}  
Here, $\rho$, $e$, and the $v_{i}$ are the mass density, internal energy density, 
and velocity at each mesh point and $p \;= (\gamma-1) e$ and {\bf{Q}} are the 
gas pressure and the von Neumann-Richtmeyer artificial viscosity tensor \citep{
ns92}. ZEUS-MP evolves these equations with a second-order accurate monotonic 
advection scheme \citep{vl77} in one, two, or three dimensions on Cartesian (XYZ), 
cylindrical (ZRP), or spherical polar (RTP) coordinate meshes.  Our augmented 
version of the publically-available code self-consistently couples primordial gas
chemistry \citep{wn06,wn08a} and multifrequency photon-conserving UV radiative 
transfer \citep{wn08b} to fluid dynamics for radiation hydroynamical transport of 
cosmological I-fronts.

\subsection{Primordial H and He Chemistry}

We evolve H, H$^{+}$, He, He$^{+}$, He$^{2+}$, H$^{-}$, H$^{+}_{2}$, H$_{2}$, and 
e$^{-}$ with nine additional continuity equations and the nonequilibrium rate 
equations of \citet{anet97}: \vspace{0.05in}
\begin{equation}
\frac{\partial \rho_{i}}{\partial t} = - \nabla \: \cdotp \; (\rho_{i} {\bf v}) 
+ \sum_{j}\sum_{k} {\beta}_{jk}(T){\rho}_{j}{\rho}_{k} - \sum_{j} {\kappa}
_{j}{\rho}_{j}, \vspace{0.05in}
\end{equation}
where ${\beta}_{jk}$ is the rate coefficient of the reaction between species j and k 
that creates (+) or destroys (-) species i, and the ${\kappa}_{j}$ are the radiative 
rate coefficients.  We assume that the species share a common velocity distribution. 
Mass and charge conservation, which are not guaranteed by either chemical or advective
updates, are enforced each time the fluid equations are solved.  The divergence terms 
and reaction network are operator-split and evolved on their respective time scales, 
as we explain in greater detail below.

Microphysical heating and cooling due to photoionization and gas chemistry is coupled 
to the gas energy density by an isochoric update that is operator-split from updates
to the fluid equations:  
\vspace{0.05in}
\begin{equation}
{\dot{e}}_{\mathrm{gas}} = \Gamma - \Lambda, \label{eqn: egas}
\vspace{0.05in}  
\end{equation}
where $\Gamma$ is the cumulative heating rate due to photons of all frequencies and 
$\Lambda$ is the sum of the cooling rates due to collisional ionization and excitation 
of H and He, recombinations of H and He, inverse Compton scattering (IC) off the CMB, 
bremsstrahlung emission, and H$_2$ cooling \citep{gp98}.

\subsection{Radiative Transfer}

Our photon-conserving UV transport \citep{anm99,mel06}, which is distinct from the 
flux-limited diffusion native to the public release of ZEUS-MP, solves the static 
approximation to the equation of transfer in flux form to compute radiative rate 
coefficients for the reaction network at every point on the coordinate mesh \citep{
anm99}. As currently implemented, our code can transport photons from a point source 
centered in a spherical grid or in plane waves along the $x$ or $z$-axes of Cartesian 
or cylindrical meshes.  In our models, the radiation from the star is treated as a 
plane wave that propagates along the positive $z$-axis.  Plane waves are a good 
approximation to the UV ionizing flux at the interhalo separations in our study, but 
we attenuate their intensity by 1/$R^2$ to account for geometrical dilution.

As in \citet{wet08b}, we discretize the blackbody photon emission rates of the stars
in our survey with 40 uniform bins from 0.755 to 13.6 eV and 80 logarithmically spaced 
bins from 13.6 eV to 90 eV, again normalizing them by the total ionizing photon rates 
for Pop III stars by \citet{s02}.  The radiative reactions in our models are listed 
in Table 1 of \citet{wn08b}.  We do not evaluate H$_2$ photodissociation rates with 
radiative transfer.  Instead, we calculate them along rays parallel to the direction
of radiation flow using self-shielding functions modified for thermal broadening as
prescribed by \citet{db96} to approximate the effects of gas motion.  They are shown 
in equations 9 and 10 of \citet{wn08b}.

\subsection{Radiation Forces}  

Since our prior survey of radiative feedback, we have implemented momentum deposition 
in the gas due to ionizations.  Radiation pressure in ionizing UV transport comes into 
play at two locations: at the I-front itself and in recombining gas in the H II region.
\citet{wn06} examined the acceleration of fluid elements at the front and found that it 
was large but momentary, and that its inclusion alters the velocity of the front by only 
1 - 2 km s$^{-1}$.  Direct momentum deposition within the H II region is only prominent 
where gas is very dense, like the center of a cosmological minihalo being evaporated by
a star at its center. There, rapid successive cycles of ionization and recombination can 
impart radiation forces to the gas that are hundreds of times the strength of gravity at 
early times \citep[lower left panel of Figure 1 in][]{ket04}. As these forces propel gas 
near the center of the halo out into the H II region, its densities and recombination 
rates fall, so more ionizing photons from the star reach the I-front.  This higher flux 
results in I-fronts that are faster than when such forces are not included.  Thus, in 
early UV breakout, radiation forces should speed up the I-front, and in trial runs we 
find that D-type fronts are 10 - 20\% faster than when momentum transfer due ionizations 
is neglected. However, this effect is transient: after the internal rearrangement of gas 
deep within the H II region dilutes its interior, radiation forces there sharply fall. 

We expect much weaker forces in this study because the front climbs a density gradient 
as it approaches the halo rather than descending one, so the gas behind the I-front is
always relatively diffuse.  Since new ionizations due to recombinations are far less 
frequent, less momentum will be imparted to the gas, so we include these effects only 
for completeness.  Updates to the gas velocities are straightforward since momentum
deposition due to direct photons from the source is always parallel to the direction 
of radiation flow. We describe the time scales on which momentum updates are performed 
below.

\subsection{Adaptive Subcycling}

A hierarchy of highly disparate characteristic time scales arises when gas dynamics, 
radiative transfer, and primordial chemistry are solved in a given application.  The 
three governing times are the Courant time, the chemical time
\begin{equation}
t_{chem} = 0.1 \, \displaystyle\frac{n_{e} + 0.001 n_{H}}{{\dot{n}}_{e}},
\end{equation}
and the photoheating/cooling time \vspace{0.05in}
\begin{equation}
t_{hc} = 0.1 \displaystyle\frac{e_{gas}}{{\dot{e}}_{ht/cool}}.\vspace{0.05in}
\end{equation}
Their relative magnitudes can seamlessly evolve throughout a single application. 
For example, when an I-front propagates through a medium, photoheating times are 
often smaller than Courant times, and chemical time scales are usually shorter than 
either one. On the other hand, fossil H II regions can cool faster than they 
recombine, so cooling times become shorter than chemical times. The key to solving 
all three processes self-consistently is to evolve each on its own timescale without 
restricting the entire algorithm to the shortest of the times. To successfully deal 
with both I-fronts and relic H II regions, an algorithm must adaptively reshuffle 
the time scales on which the three processes are solved. Implicit schemes are 
sometimes applied to stiff sets of differential equations like those in our 
model because they are unconditionally stable over the Courant time.  However, 
accurate I-front transport in stratified media often requires restricting updates 
to both the gas energy and fluid equations to photoheating times in order to capture 
the correct energy deposition into the gas, and linear systems solves over such short 
time scales would be prohibitive in more than one or two dimensions.  Enforcing photon 
conservation in implicit schemes can also be problematic.  

\begin{figure*}
\epsscale{1.15}
\plottwo{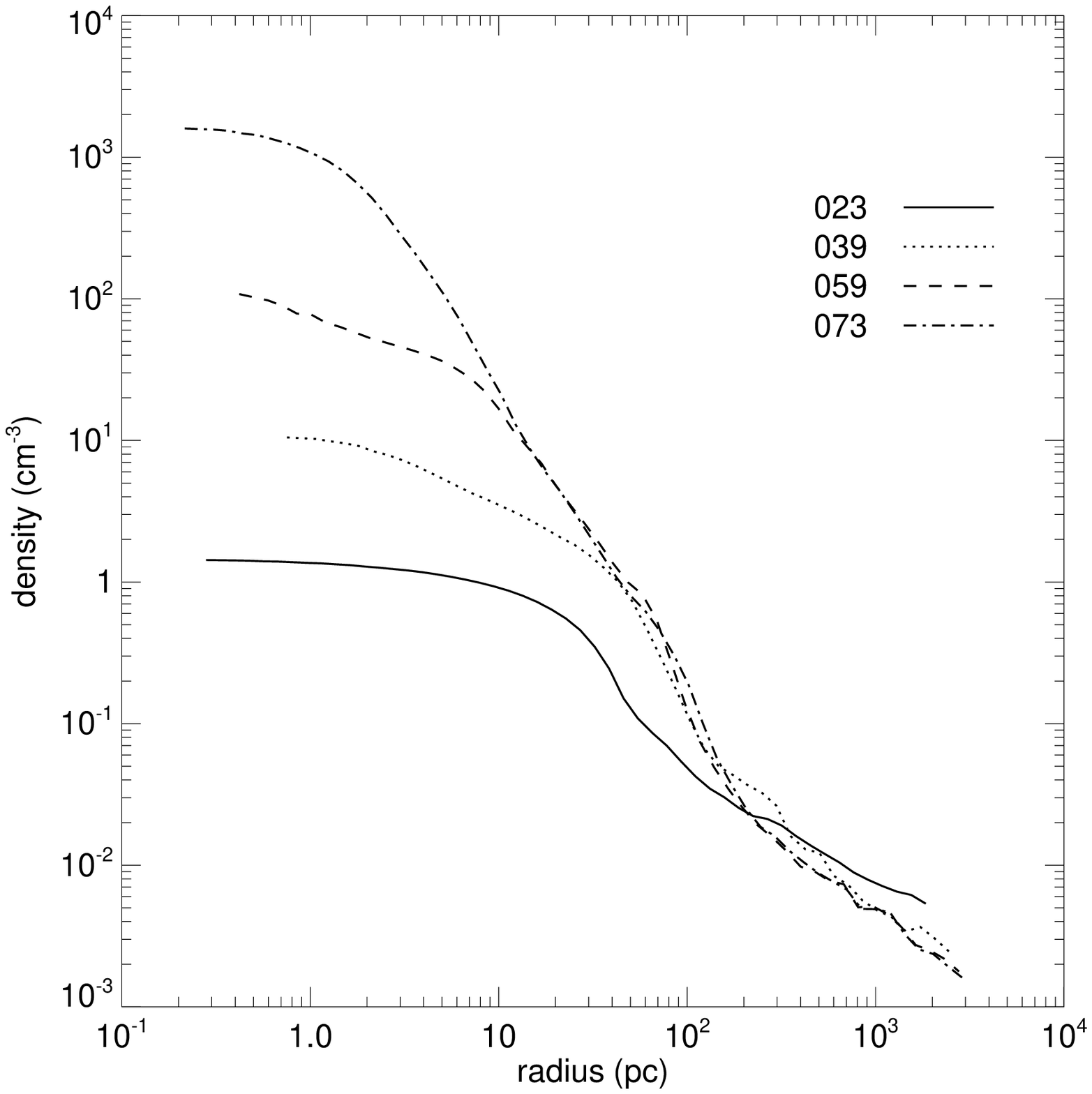}{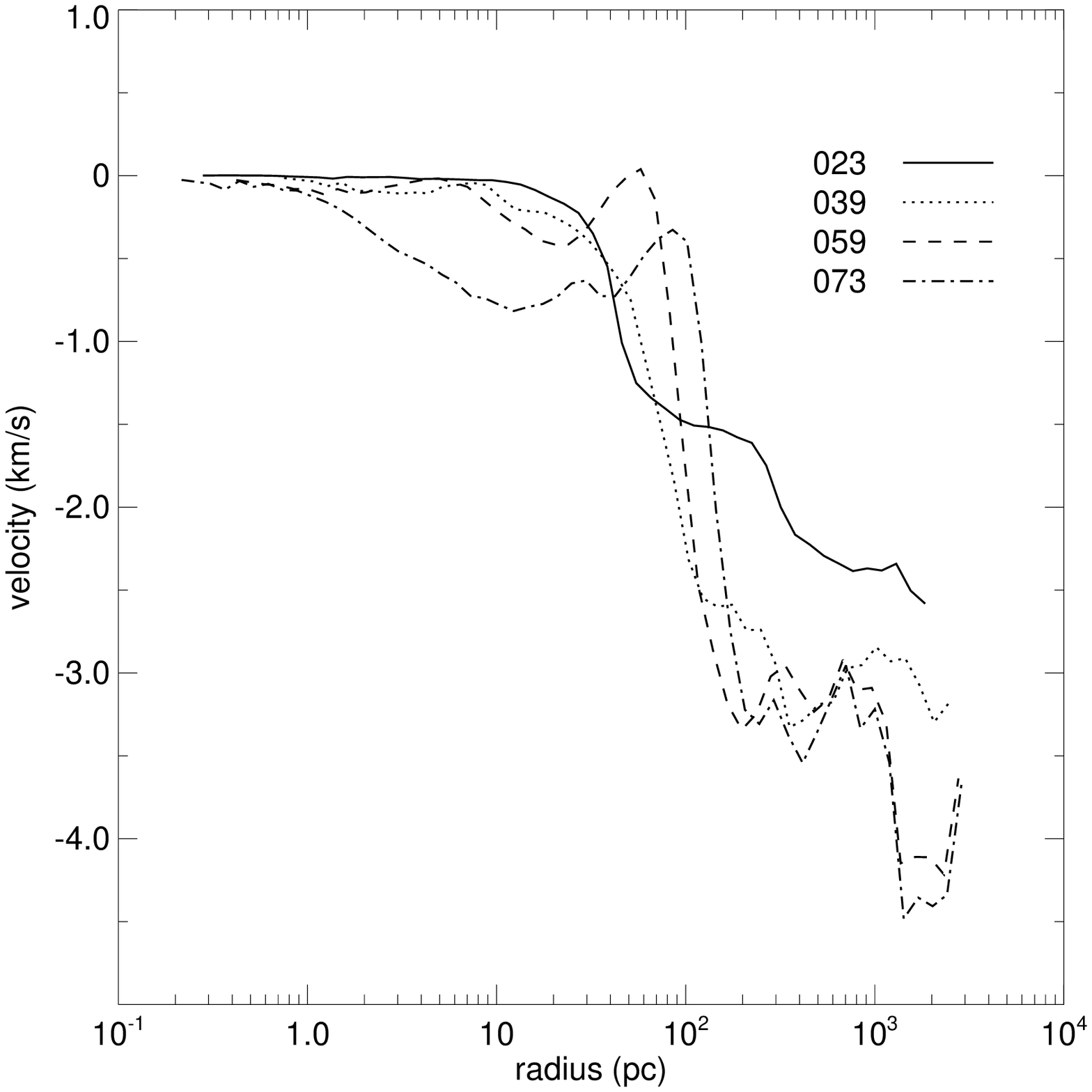}\vspace{0.15in}
\caption{Spherically-averaged baryon profiles for the 1.35 $\times$ 10$^5$ \Ms \ halo
at four stages of evolution. The redshifts of the 023, 039, 059, and 073 profiles are 
23.9, 17.7, 15.6, and 15.0, respectively, with corresponding central densities of 1.43, 
10.5, 108, and 1596 cm$^{-3}$.  Left: densities.  Right: radial velocities. 
\label{fig:halop}} 
\vspace{0.1in}
\end{figure*}

We instead subcycle chemical species and gas energy updates over the minimum of the 
chemical and heating/cooling times until the larger of the two has been crossed, at 
which point we perform full hydrodynamical updates of gas densities, energies, and 
velocities.  These times are global minima for the entire grid.  The chemical times 
are defined in terms of electron flow to accommodate all chemical processes rather 
than just ionizations or recombinations.  Adopting the minimum of the two times for 
chemistry and gas energy updates enforces accuracy in the reaction network when $
t_{chem}$ becomes greater than $t_{hc}$, such as in relic H II regions.  Our 
adaptive subcycling scheme is described in greater detail in \citep{wn08a}.

\section{Halo Models}

The $1.35 \times 10^5$~M$_\odot$ halos we study are computed from cosmological initial
conditions in the Enzo adaptive mesh refinement (AMR) code as described in detail in 
\citet{wet08b}.  The halo baryon density and temperature profiles were spherically 
averaged and then mapped onto a two-dimensional cylindrical coordinate (ZR) grid in 
ZEUS-MP. The assumption of radial symmetry in two dimensions is approximate but valid, 
given the spheroidal nature of the halos.  A single, symmetrized baryon profile better 
represents all halos of this mass than the three-dimensional profile of just this halo
because small differences in angular distribution of the baryons are averaged out. We 
center the halo on the $z-$axis so that only its upper hemisphere resides on the grid.  
The mesh boundaries are -125 pc and 125 pc in $z$ and 0.01 pc and 125 pc in $r$.  The 
grid is 1000 zones in $z$ and 500 zones in $r$ with a spatial resolution of 0.25 pc.  
Outflow conditions are assigned to the upper and lower boundaries in $z$ and reflecting 
and outflow conditions are imposed on the inner and outer boundaries in $r$, respectively.
  
Dark matter gravity is included by constructing a separate potential that cancels 
pressure forces everywhere on the mesh, thus imposing hydrostatic equilibrium on the 
halo, and then holding the potential fixed throughout the simulation.  Force updates to 
gas velocities are performed with the potential every hydrodynamical time step during the 
simulation.  Neglecting dark matter dynamics introduces no serious errors because the gas 
in the halo evolves on much shorter time scales than either the Hubble time or merger times, 
which at $z \sim$ 20 are approximately 20 Myr.  Updates to the self-gravity of the gas are 
also performed every hydrodynamical time step by evolving Poisson's equation with a 
two-dimensional conjugate gradient (CG) solver.    

In reality, the halo does have infall velocities as shown in Fig \ref{fig:halop}, but 
they are minor.  Their inclusion would only slightly enhance core densities during 
photoevaporation.  The composition of the gas in our models was primordial, 76\% H and 
24\% He by mass.  We assume ionized and H$_2$ fractions of 1.0 $\times$ 10$^{-4}$ and 
2 $\times$ 10$^{-6}$, respectively.  These values are consistent with remnant free 
electron fractions from the era of recombination at the redshift $z =$ 20 we take for
our models and with those expected from collisional ionizations in cosmological 
accretion shocks.  We consider the four evolutionary stages of the halo in our earlier 
work, corresponding to central densities $n_c$ = 1.43, 10.5, 108, and 1596 cm$^{-3}$, 
respectively.  Each halo profile is illuminated at 150, 250, 500, and 1000 pc, which 
are typical of interhalo separations in a cluster.  Each profile is illuminated by 25, 
40, 60 and 80 \Ms \ stars for their respective lifetimes, 6.46, 3.86, 3.46 and 3.01
Myr, for a total of 64 models.  These stars sample the lower end of the Population III 
mass spectrum.  After the death of the star the halo is further evolved to a total of 
10 Myr, or about half of a merger time, to determine the effect of relic H II region 
dynamics on the gas in the core.

\subsection{Criteria for Star Formation} \label{sect:sf}

Unlike AMR or Lagrangian SPH algorithms, the fixed Eulerian 
grid in ZEUS-MP lacks the resolution to follow the collapse of the baryons in the halo 
into a new star.  However, numerical simulations demonstrate that primordial star 
formation is a very robust process:  if a halo is in a state in which its baryons can 
form a star, it will within a few Myr.  Since our halo is one in which star formation 
has already been proven to occur, if the final density and H$_2$ fraction of its core 
is greater than or equal to its pre-illumination levels, a star will form at its center.  
In some cases the relic I-front shock passes through the center of the halo and imparts 
an impulse to the gas there.  However, if the final velocity of the perturbed baryons 
within a parsec of the core is less than the escape speed from the halo, they will be 
trapped by the dark matter potential and still form a star.  We therefore adopt the 
simple criteria that if central baryon densities and H$_2$ fractions in the halo are 
the same or greater than those just before illumination, and if central gas velocities 
are lower than the escape speed, a star will form in the halo.  If we approximate the 
binding energy $E_{\mathrm{B}}$ of the baryons to the halo by that of a homogeneous 
sphere, \vspace{0.1in}
\begin{equation}
E_{\mathrm{B}} = \frac{3}{5} \; \frac{GM_{\mathrm{h}}m_{\mathrm{b}}}{R_{\mathrm{vir}}},\vspace{0.1in}
\end{equation}
where $M_{\mathrm{h}}$, $m_{\mathrm{b}}$, and $R_{\mathrm{vir}}$ are the total halo 
mass, baryon mass, and virial radius of the halo, respectively, then the escape speed 
$v_{\mathrm{esc}}$ is
\begin{equation}
v_{\mathrm{esc}} \sim \left(\displaystyle\frac{6}{5} \; \frac{GM_{\mathrm{h}}m_{\mathrm{b}}}{R_{\mathrm{vir}}}\right)^{\frac{1}{2}}.
\end{equation}
For the halo in our models, $v_{\mathrm{esc}} \sim$ 2.6 km s$^{-1}$.

\section{Results}

\begin{figure}
\resizebox{3.45in}{!}{\includegraphics{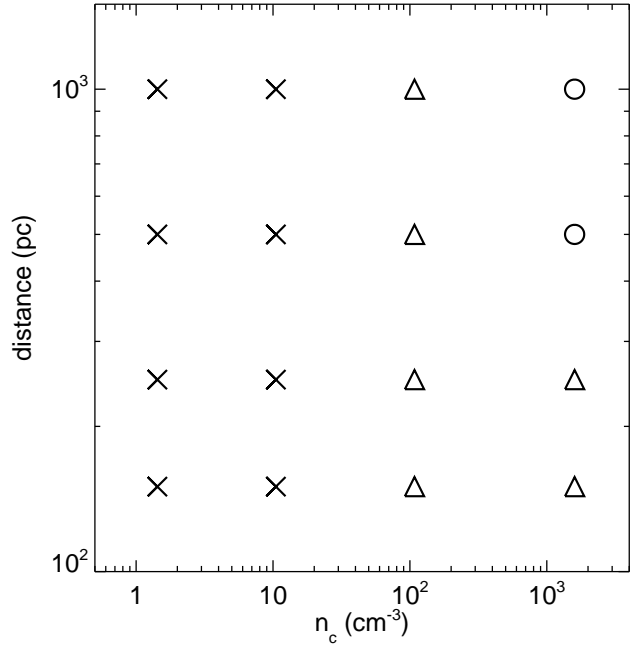}}
\vspace{0.1in}
\caption{Star formation in a 1.35 $\times$ 10$^5$ \Ms \ halo in the vicinity
of a 120 \Ms \ star, from \citet{wet08b}.  The central baryon densities $n_c$ 
of the halo at the time of illumination were 1.43, 10.5, 108, and 1596 cm$^{-1}$, 
respectively.  The 120 \Ms \ star was at 150, 250, 500 and 1000 pc. Completely 
evaporated halos with no star formation are labeled by crosses and halos with 
delayed or undisturbed star formation are represented by triangles and circles, 
respectively. \vspace{0.1in}
\label{fig:120Ms}}  
\end{figure}

\begin{figure*}
\epsscale{1.15}
\plottwo{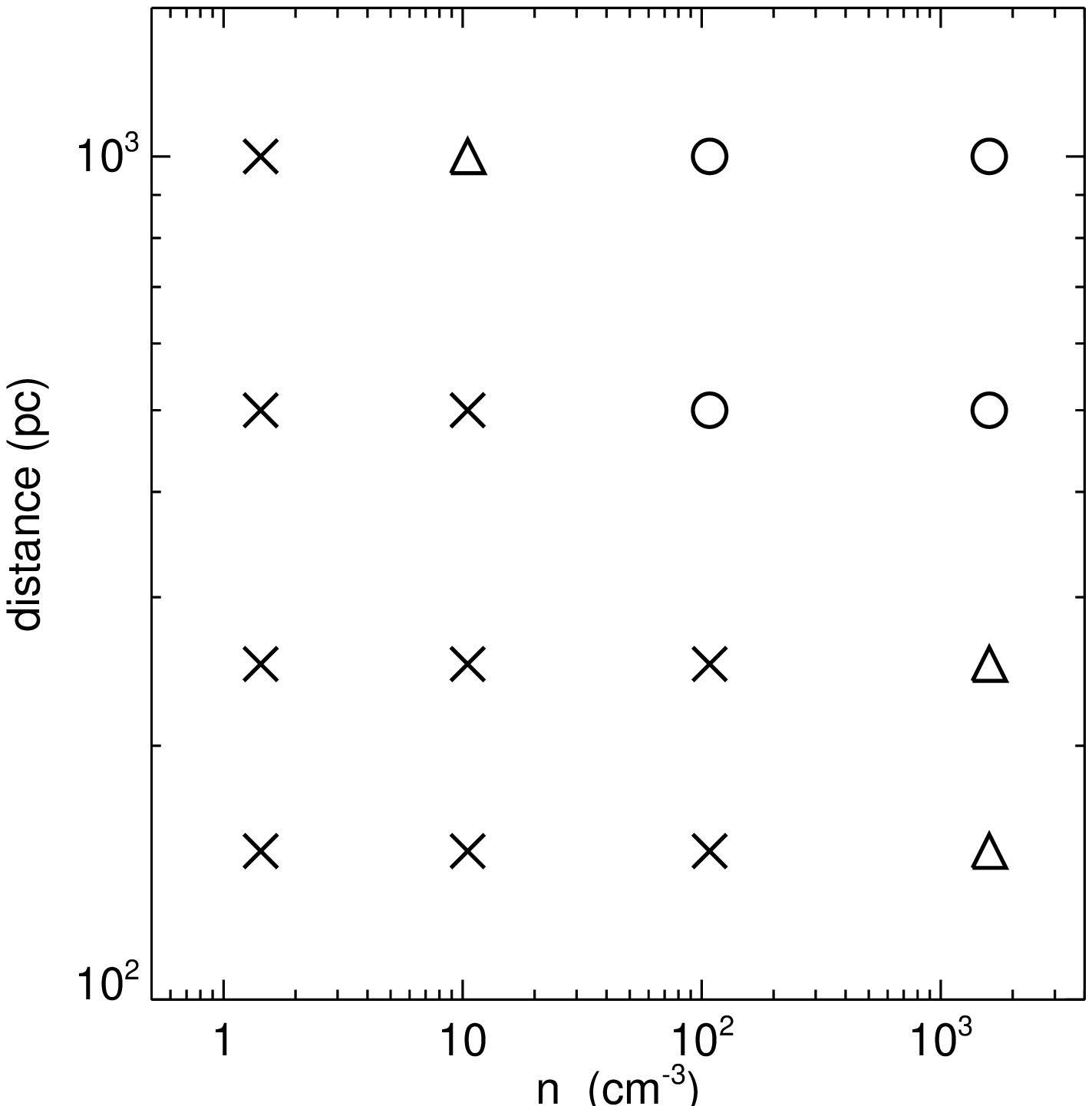}{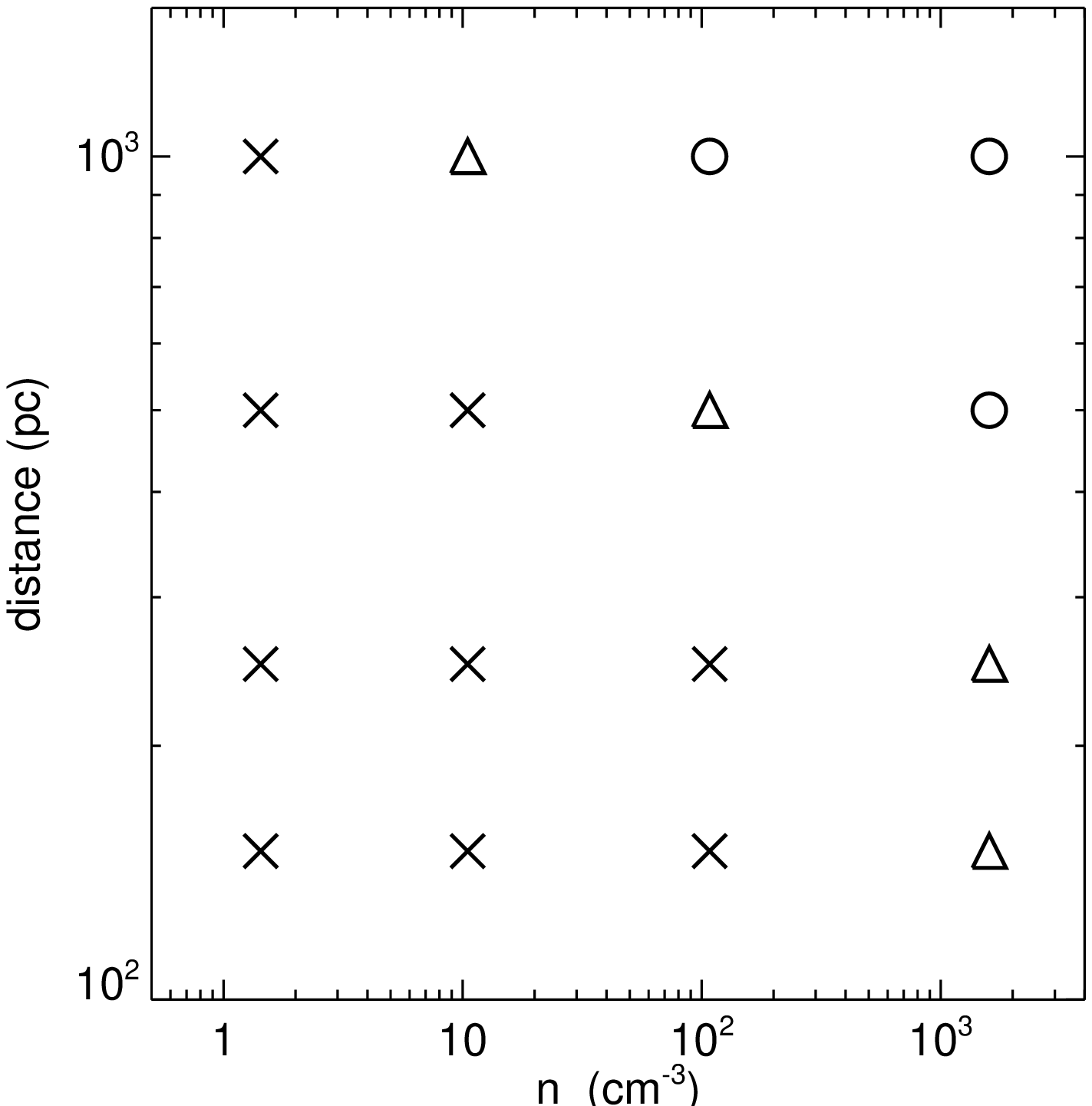}
\vspace{0.1in}
\caption{Star formation in the halo when illuminated by a 25 \Ms \ star (left) 
and a 40 \Ms \ star (right) at four central baryon densities $n_c$, 1.43, 10.5, 
108, and 1596 cm$^{-3}$, and at four distances from the star, 150, 250, 500 and 
1000 pc, which are typical interhalo spacings within a cluster.  Completely 
evaporated halos with no star formation are labeled by crosses, and halos with 
delayed or undisturbed star formation are indicated by triangles and circles, 
respectively.  
\label{fig:25Ms}} 
\end{figure*}

\begin{figure*}
\epsscale{1.15}
\plottwo{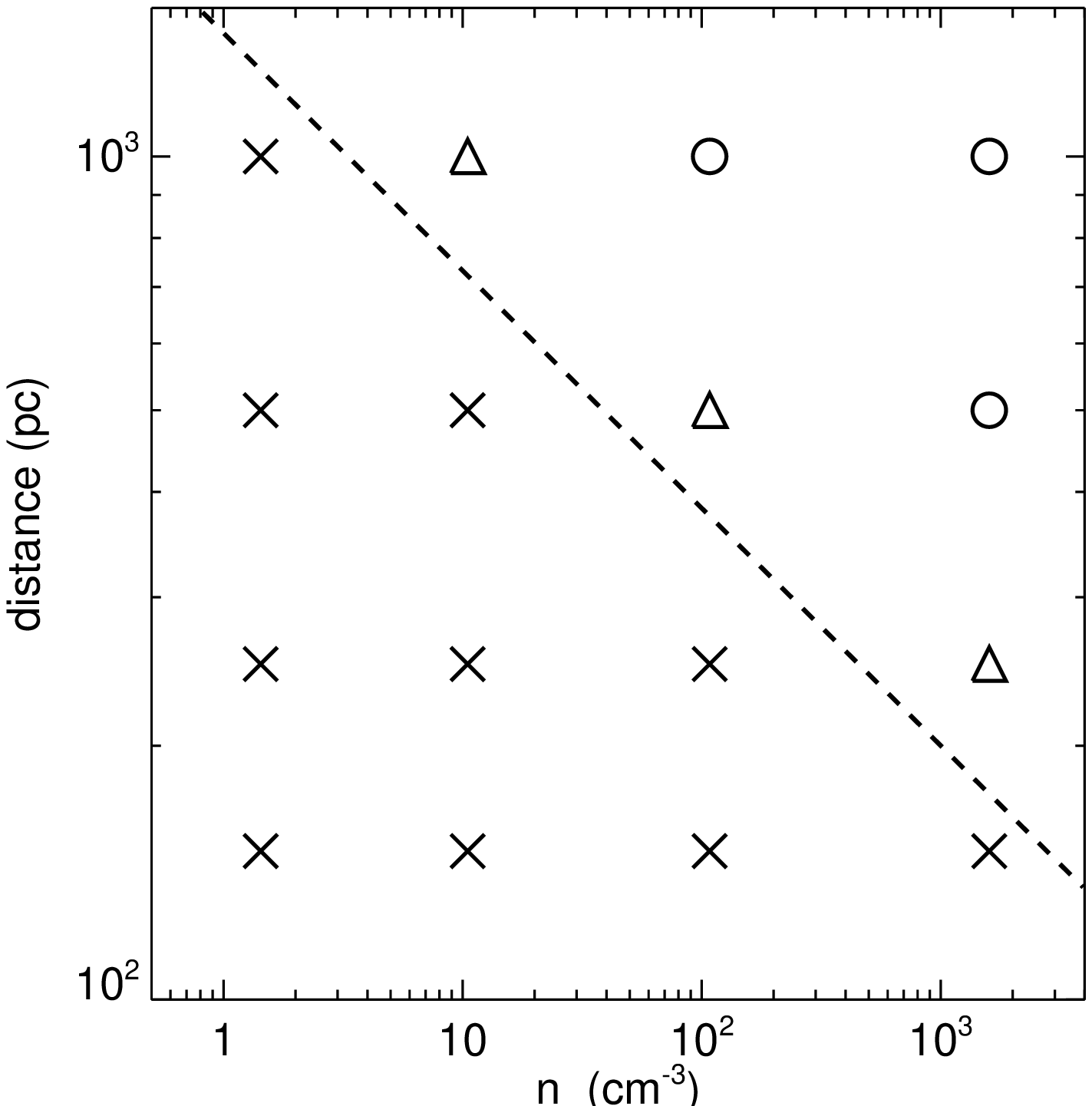}{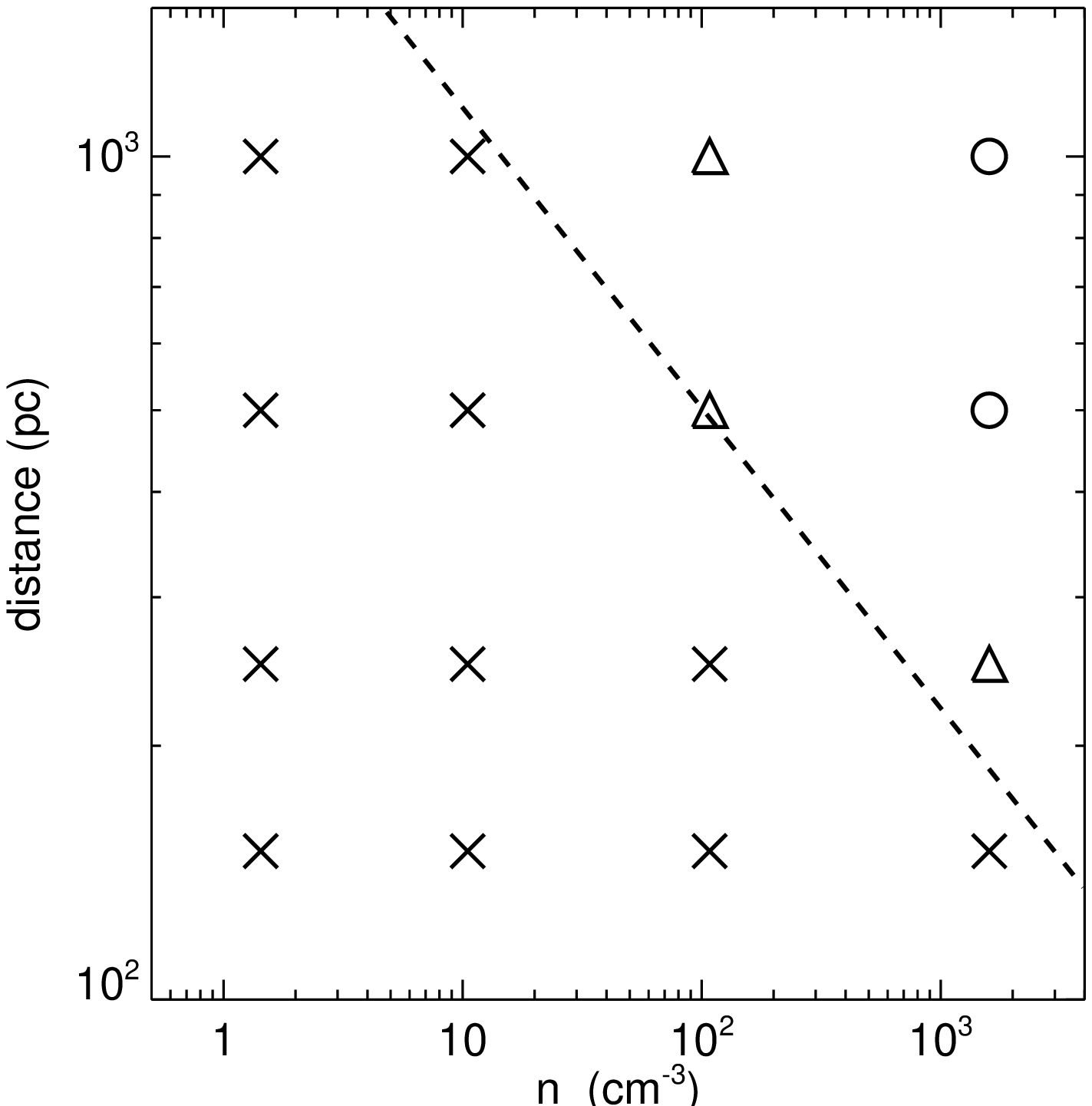}
\vspace{0.1in}
\caption{Star formation in the halo in the vicinity of a 60 \Ms \ star (left) 
and an 80 \Ms \ star (right) at the central gas densities and distances to the 
star considered in our study.  Completely ionized halos with no star formation 
are labeled by crosses, and halos with delayed or undisturbed star formation 
are indicated by triangles and circles, respectively.  The dotted lines mark
the threshold for star formation in the evaporated halos, above which it 
proceeds and below which it is quenched.  \vspace{0.1in}
\label{fig:60Ms}} 
\end{figure*}

The \citep{wet08b} study found several possible final fates for halos
photoevaporated by a 120 \Ms \ star.  If the halo is very diffuse, with 
$n_c < 1$ cm$^{-3}$, it is completely evaporated anywhere it resides in 
the cluster, with no star formation.  If the halo is more evolved, with 
$n_c > 1000$ cm$^{-3}$, its core is shielded from both ionizing and LW 
flux and star formation proceeds there without delay. In halos of medium 
central density the relic ionization front shock either compresses and 
enriches the core with H$_2$, potentially hastening its cooling and 
collapse into a star, or its residual momentum disrupts the core, 
delaying or suppressing star formation.  For comparison to our 25 - 80
\Ms \ models, which are evolved to 10 Myr, we ran the 120 \Ms \ models
of \citep{wet08b} out to the same time. We list final outcomes for star 
formation in these simulations according to the criteria set forth in
section \ref{sect:sf} in Figure \ref{fig:120Ms}.

\subsection{25 - 80 \Ms \ Stars}

In Figures \ref{fig:25Ms} and \ref{fig:60Ms} we tabulate the effects 
of local ionizing and LW radiation on star formation in the halo near 
25, 40, 60 and 80 \Ms \ stars.  First, we note that in each of these 
stars new star formation obeys the same trends as in a 120 \Ms \ star:  
it is first suppressed, then delayed, and finally unaffected as its 
central baryon density and distance to the star grow.  These trends 
are primarily due to the momentum with which the I-front shock reaches 
the core of the halo.  As explained earlier, when star formation is 
delayed it is due either to disruption of the center of the halo by 
the relic I-front shock, LW photodissociation of the core of the halo, 
or both.  If star formation is unaffected it is because the core is 
completely shielded from LW flux, which also guarantees that the relic 
I-front shock will not reach the core before it forms a star. From the 
upper left to lower right in the 60 and 80 \Ms \ panels runs a line 
that marks the boundary between quenched star formation and delayed or 
undisturbed star formation (corresponding lines for 25 and 40 \Ms \ 
stars, which die in SN explosions whose kinetic feedback must also be 
taken into account, are shown in section 4.2).  The suppression of new 
star formation in the models lying just below this line is usually 
unambiguous: the shock completely sweeps baryons from the core of the 
halo.  Only in models lying just above this line does star formation 
vary with the mass of the illuminating star.  The line advances 
gradually up and to the right with neighbor star mass, but the shift 
is minor from 25 - 80 M$_{\odot}$.  In a halo at a fixed central 
baryon density and distance from the star, the formation of a new star
is at most delayed if it was once unaffected, or halted if it was once 
delayed, as the mass of the illuminating star varies from 25 - 80 M$_{
\odot}$.  Star formation remains unchanged above and below this narrow 
band. We note that the lines are drawn to cut the space between symbols 
evenly and as such should be taken as order of magnitude estimates. 

\begin{figure*}
\epsscale{1.17}
\plottwo{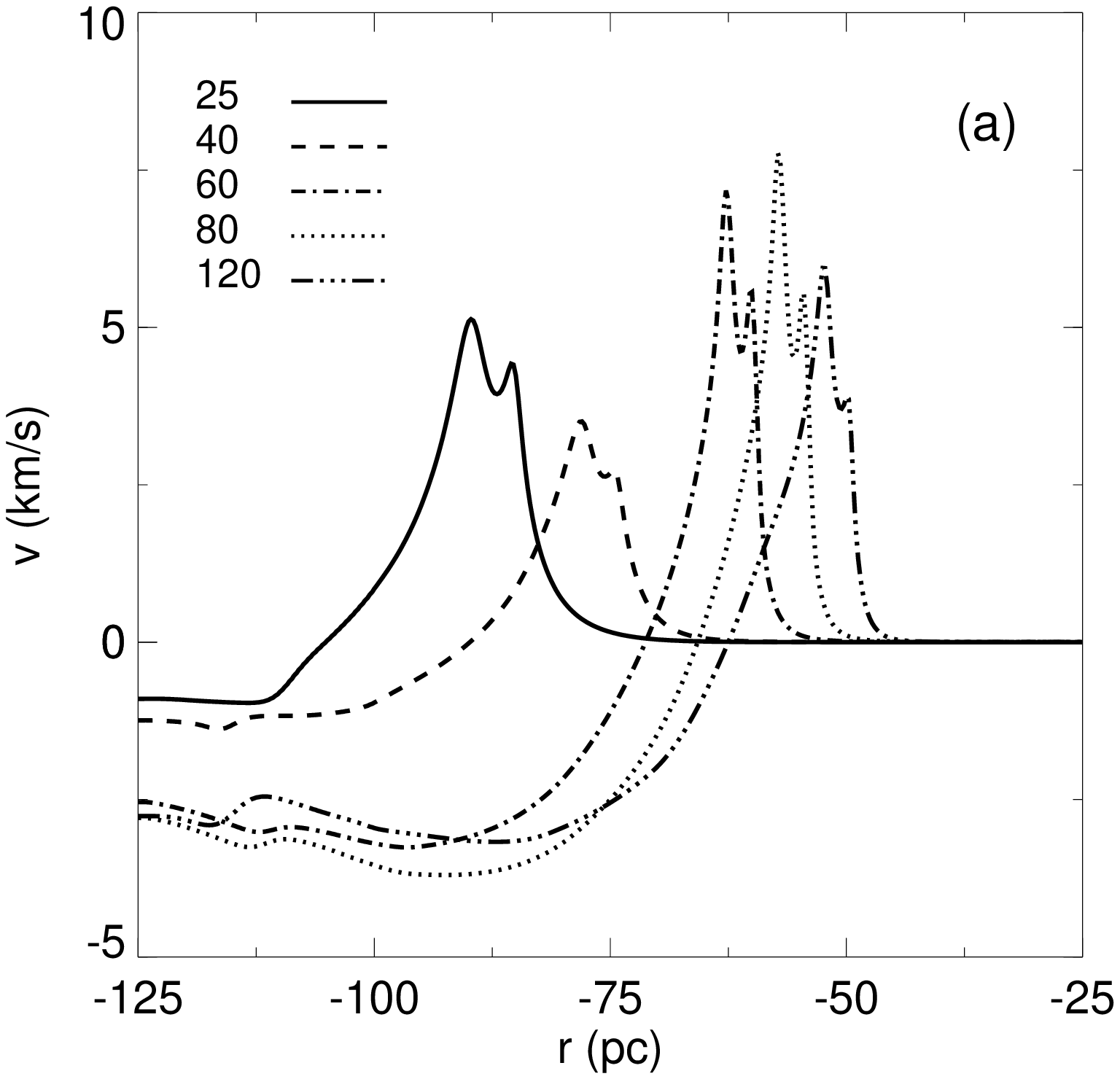}{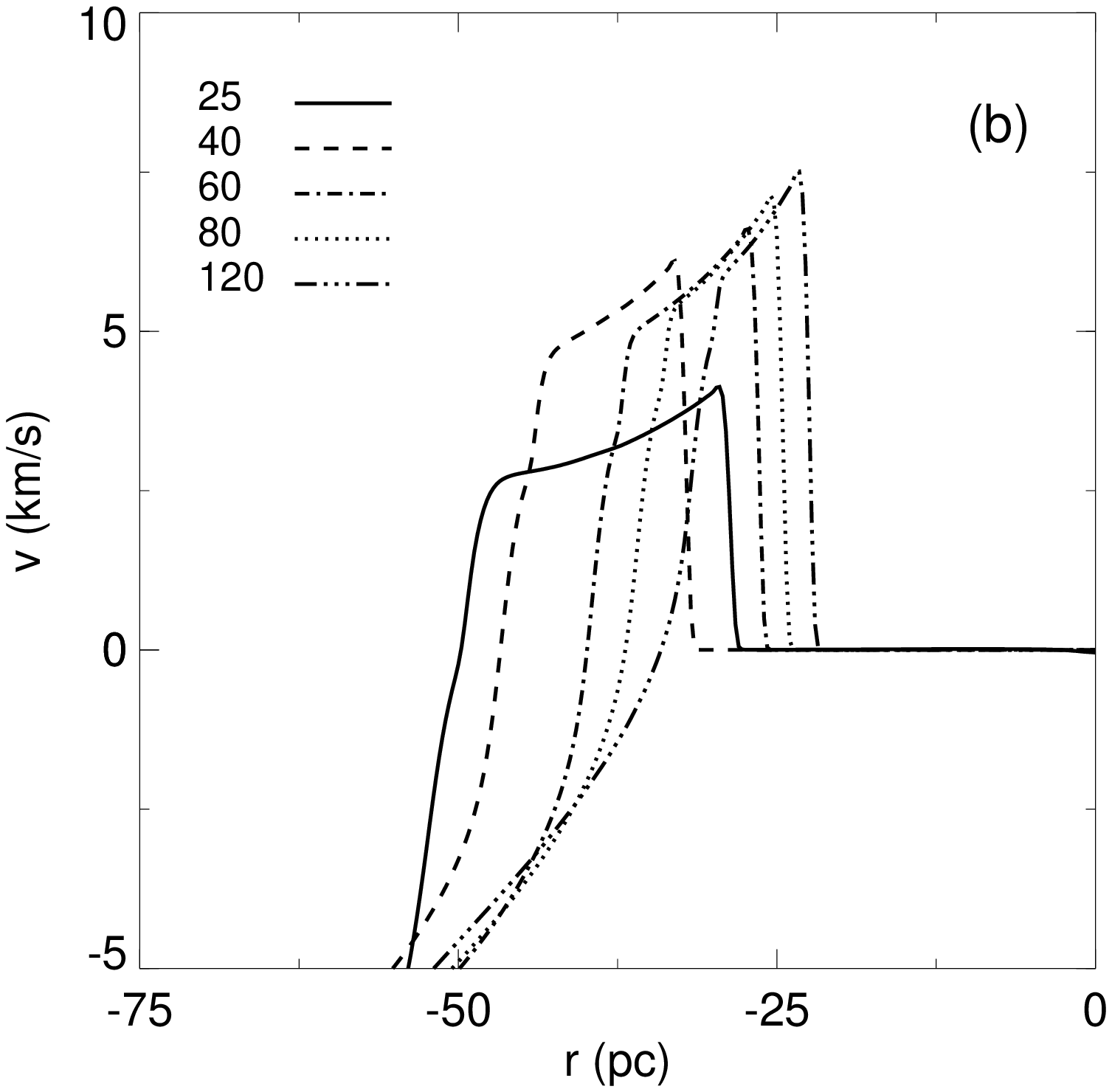}
\caption{I-front gas velocity profiles through the central axis of the
halo for 25, 40, 60, 80, and 120 \Ms \ stars 500 pc from halo 059 ($n_c$ 
= 108 cm$^{-3}$). Panel (a): velocity at the time each front transforms 
from R-type to D-type. Panel (b): gas velocity profiles at the time each 
star dies.  
\label{fig:vel}} 
\end{figure*}

Why is star formation in the satellite halo relatively uniform with 
neighbor star mass?  On one hand, the dimmer flux of low-mass Pop III 
stars causes the front to transform from R-type to D-type further out 
from the center of the halo. We plot velocity profiles for the I-fronts 
at the moment they become D-type in the $n_c =$ 108 cm$^{-3}$ halo 500 
pc from 25 - 120 \Ms \ stars in panel (a) of Figure \ref{fig:vel}.  In
each profile the transition distance from the core is marked by the 
position of the forward peak, which is at 80, 75, 60, 55 and 50 pc for 
the 25, 40, 60, 80 and 120 \Ms \ stars, respectively. On the other hand, 
lower-mass stars are also longer lived, so they drive the I-front into 
the halo for greater times.  This collapses the dispersion in I-front 
positions in the halo to 10 pc by the time the star dies, as shown in 
panel (b) of Figure \ref{fig:vel}.  In this halo, the I-fronts reach 
the core with nearly the same velocity, which is well below the escape 
speed, with each star.  This pattern holds for all halos in which the 
I-front falls short of the core when the star dies, when star formation 
is delayed or uninterrupted.  Note that at early times the velocity
profile is split into two smaller peaks.  This is due to penetration
of hard UV photons into the dense shocked gas ahead of the front, which
drives a backflow in the frame of the shock.  This is a common feature
of I-fronts due to hard UV spectra, as discussed in detail in section
4.1 and Figure 17 of \citet{paper1}. The evolution in spectral profile 
from 25 - 120 M$_{\odot}$, which causes ionized gas temperatures to 
rise by more than 50\% in the H II region, accounts for the variation 
in peak velocity in Figure \ref{fig:vel}.  Had each of these I-fronts 
been driven by a monochromatic flux of the same magnitude and duration, 
the spread in the peaks in radius would have been even less than 10 pc.

\begin{figure}
\resizebox{3.45in}{!}{\includegraphics{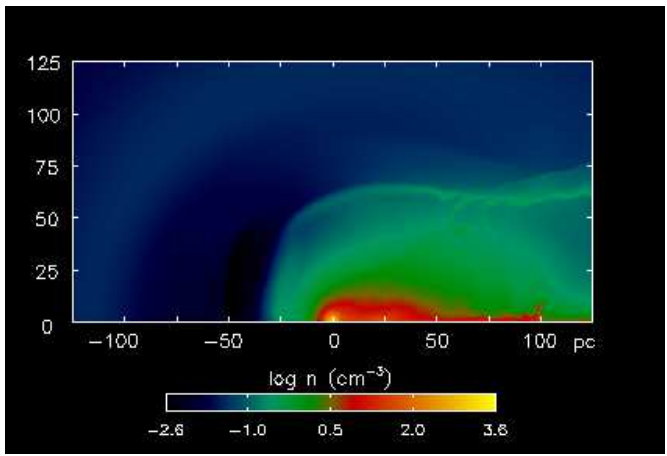}}
\vspace{0.1in}
\caption{Evaporated halo with $n_c =$ 1596 cm$^{-3}$ (halo 073) 150 
pc from a 60 \Ms \ star at 10 Myr.  The core of the halo is slightly 
displaced to the left of center by backflow from the collapsed shadow 
on the right.
\vspace{0.1in}
\label{fig:backflow}} 
\end{figure}

In the \citet{wet08b} survey, the momentum imparted by the relic I-front 
shock to the core of the halo is primarily what determines if a new star 
forms there. Here, in a few cases, the compresssion of the shadow of the 
halo toward the axis also determines if a star forms.  In our new models 
the halo forms a shadow on the same time scale as a 120 \Ms \ star but 
pressure from the surrounding relic H II region drives it inward toward 
the axis for up to twice the time before the star dies, squeezing a flow 
backward into the center of the halo. This backflow can slightly displace 
baryons from the core, as we show in Figure \ref{fig:backflow} for halo 
073 150 pc from a 60 \Ms \ star.  The clump of gas originally centered 
in the dark matter potential retains its shape and is at well below the 
escape speed of the halo. However, its slight dislocation from the center 
of the dark matter potential at 10 Myr prevents it from collapsing into 
the star that would have formed if only the relic I-front shock had been 
present.  Backflows are especially collimated in the axial symmetry we 
assume in our models but we still expect them to be present in three
dimensional halos because of their approximate sphericity.

\begin{figure*}
\epsscale{1.17}
\plottwo{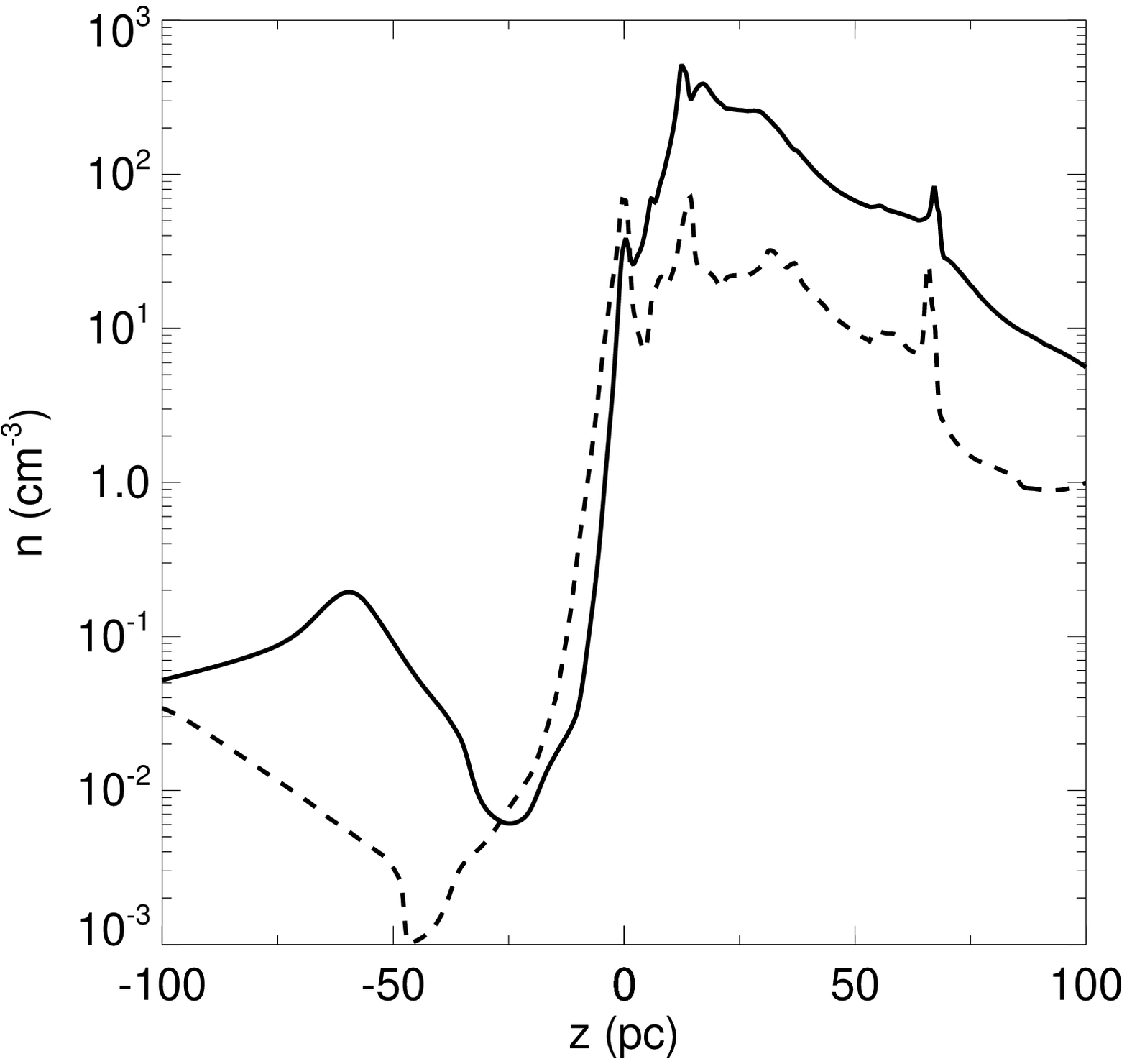}{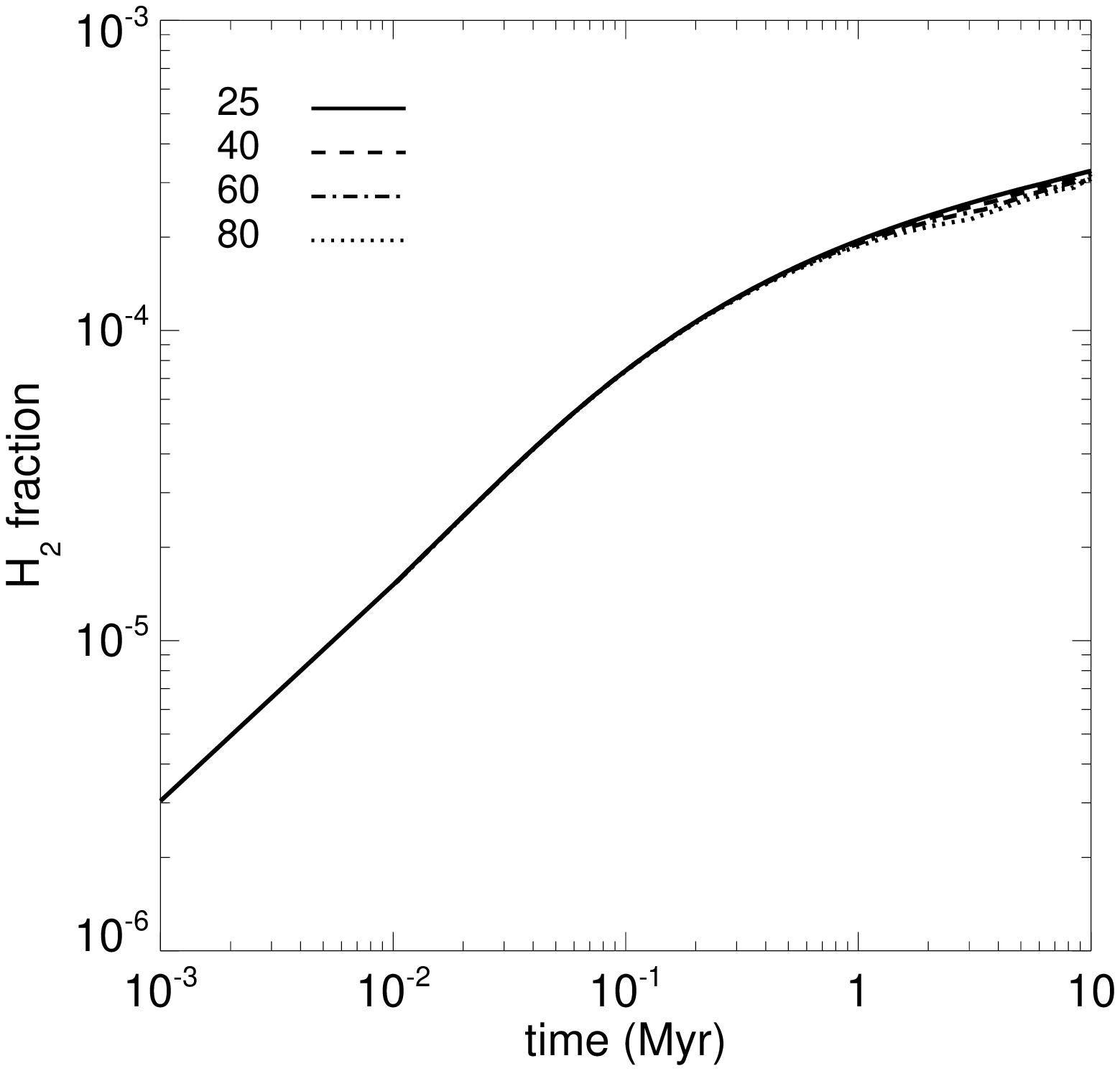}
\caption{Left panel: the flow of baryons back into the center of the 
dark matter potential from the relic H II region, as shown in these 
density profiles along the central axis of the halo.  Solid: 5.5 Myr; 
dashed: 10 Myr.  Right panel: H$_2$ mass fractions at the center of 
halo 073 500 pc from 25, 40, 60 and 80 \Ms \ stars from zero to 10 
Myr.
\label{fig:bkfl}} 
\end{figure*}

In halos where stars form without interruption during photoevaporation, 
molecular hydrogen mass fractions rapidly rise from the initialized value 
of 2 $\times 10^{-6}$ to 10$^{-4}$ when the star is turned on, even with 
LW flux, because the core self-shields, as we show for the halo at $n_c 
= $1596 cm$^{-3}$ 500 pc from 25, 40, 60, and 80 \Ms \ stars in the right 
panel of Figure \ref{fig:bkfl}.  The core collapses even as the outer 
layers of the halo are stripped away by supersonic outflows, and a star 
forms just as quickly as in the absence of radiation. Since we begin with 
cosmic mean H$_2$ fractions of 2 $\times 10^{-6}$ instead of more realistic 
values of 10$^{-4}$ for simplicity, ours are lower limits to self-shielding 
and cooling. 

One departure of the 120 \Ms \ models from the others is the delayed 
star formation at 150 and 250 pc in the halo at $n_c =$ 108 cm$^{-3}$, 
which does not occur near 25 - 80 \Ms \ stars.  This happens because 
the shock remnant traverses the core of the halo sooner, allowing gas 
to later pool in the dark matter potential and reach its original 
density at the center by 10 Myr, as shown by the dashed line in the 
left panel of Figure \ref{fig:bkfl} at 150 pc.  Such backfill allows 
a new star to form before a total of 20 Myr has elapsed, or about a 
merger time at $z \sim$ 20.  Similar flows would occur in low-mass 
runs evolved beyond 10 Myr but would probably not result in a new 
star prior to severe disruption by a merger.  The higher densities at 
earlier times in the outer regions of the halo (solid lines in Figure
\ref{fig:bkfl}) are due to compression by the relic I-front shock as 
it envelopes the halo. 
 
We find that the higher ratio of LW to ionizing photons of low-mass 
Population III stars has no effect on nearby star formation. The lower 
fluxes of these stars offset their higher ratios, and in any event LW 
photons from a single star cannot prevent a new star from forming in a 
nearby halo prior to the appearance of LW backgrounds at lower redshifts. 
They can only delay it for the life of the proximate star.  Without LW 
photons halo collapse times at $n_c \sim$ 2000 cm$^{-3}$ are 7 - 10 Myr.  
At such densities, \citet{wet08b} find that the core begins to strongly 
self-shield from local LW flux.  If the core could collapse before the 
arrival of the I-front, its $n_c$ would be far greater, be fully shielded 
from LW flux, and form a star anyway.  Halos that are completely ionized 
are too diffuse to form a star before the I-front reaches the core even 
in the absence of LW photons.  Cores that form a star after the death of 
a nearby star could not have created one sooner without LW flux because 
their star formation times are even greater than 7 Myr.  Finally, halos 
that are too severely disrupted by the relic I-front shock to form a star 
cannot collapse prior to the arrival of the shock in the absence of LW 
radiation for the same reason.  Thus, in none of these outcomes would a 
star have formed if there had been no LW flux.  In the latter two cases, 
we find that H$_2$ fractions at the center of the halo reacquire their 
original values 100 - 200 kyr after the star dies and that cooling and 
collapse of the core begins anew, augmented in some cases by molecular 
hydrogen advected into it by the relic I-front shock.  Any suppression 
or delay of star formation is entirely due to bulk flows driven into 
the core driven by relic I-front shock and shadow dynamics, not to the 
destruction of H$_2$, which quickly reconstitutes in the core after the 
star dies.  Thus, local ionizing UV flux governs new star formation in 
clustered halos, not local LW photons.

An exception to this is when the baryons are very close to the star, 25 - 
150 pc.  In such cases, when a star is irradiating a clump of baryons in 
the same halo, \citet{hus09} find that clouds with free-fall times that
are shorter than ionization times can be prevented from collapsing by 25 
\Ms \ stars because they have the highest LW/ionizing photon ratio.  This
never occurs in our models because of the much lower LW fluxes at typical 
halo distances within the cluster, but such scenarios are quite relevant 
to the formation of Population III binaries within a halo \citep{turk09}. 
Persistent LW backgrounds at lower redshifts \textit{can} by themselves 
prevent secondary star formation in the cluster by not allowing partially 
evaporated cores to cool after the death of the star.  However, the 
exclusion of radiation hydrodynamical effects in past studies may have 
led them to overestimate this effect, as we discuss in $\S$ 5. 

\subsection{Kinetic Feedback by SNe}

Unlike the 100 - 120 \Ms \ stars assumed in most local radiative feedback 
studies, 25 - 50 \Ms \ stars die in SN explosions \citep{hw02,Tominaga2007}. 
Their remnants can ram-pressure strip baryons from nearby halos and prevent 
them from forming a star when one otherwise would have been created. \citet{
ss09} examined kinetic feedback by SNe on star formation in satellite halos 
with semi-analytical arguments that neglected radiative preprocessing of the 
halo by the progenitor star. They found that SNe generally preempt any star 
formation that fails to occur prior to the arrival of the remnant at the halo. 
\citet{cr08} simulated the interaction of SN remnants with much more massive 
halos at lower redshifts to assess the degree to which metals become mixed 
with gas deep in their interiors.  These calculations, which also excluded 
prior photoevaporation by UV backgrounds, found some mixing in the outer 
layers of the halos due to Kelvin-Helmholtz instabilities but that no metals 
reached their interiors.  These models did not address SNe feedback on star 
formation because the potential wells of the halos were much deeper than 
those in which the first stars formed, so the remnants could not strip gas 
from their cores.  Detailed numerical models of the collision of primordial 
supernova remnants with the relic H II regions enveloping partially exposed 
cores are needed to establish the actual fate of star formation in these 
circumstances.

\begin{figure}
\vspace{-0.38in}
\resizebox{3.45in}{!}{\includegraphics{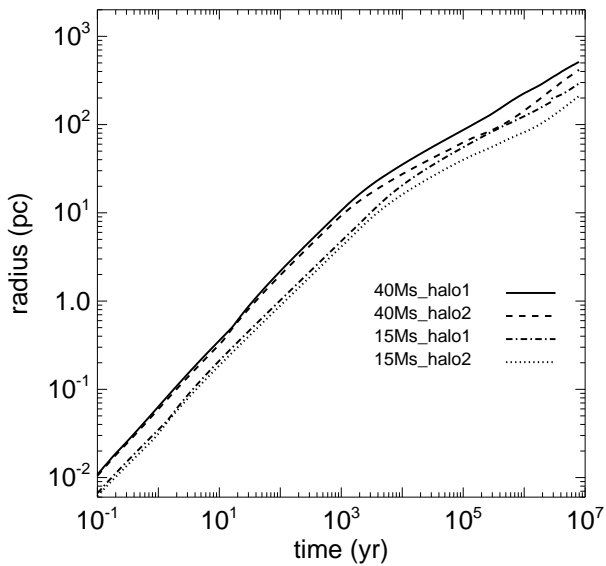}}
\caption{Radii of 15 \Ms \ supernova and 40 \Ms \ hypernova remnants in 6.9 
$\times$ 10$^5$ \Ms \ and 2.1 $\times$ 10$^6$ \Ms \ halos (halo1 and halo2, 
respectively) from \citet{wet08a}.\vspace{0.1in}
\label{fig:rvst}}  
\end{figure}

Supernovae are not invariably fatal to local star formation because the time 
scales on which the shock reaches outlying halos may be longer than those on 
which baryons collapse within them, and in some cases the remnant never 
reaches them. \citet{get07} found that SN remnants typically expand to only 
half the radius of the relic H II region of the progenitor because they come 
into pressure equilibrium relatively quickly in the warm recombining gas. 
\citet{wet08a} computed the radius of the remnant as a function of time for 
core collapse SNe, hypernovae, and pair-instability SNe (PISN) in the fossil 
H II regions of 10$^5$ - 10$^7$ M$_{\odot}$.  In their simulations, ejecta 
from the explosions of 15 - 40 \Ms \ stars propagate at most 400 - 500 pc in 
10 Myr.  We plot the radii of 15 and 40 \Ms \ SN remnants in those models in 
Figure \ref{fig:rvst}.  As shown in Figure 10 of \citet{wet08a}, the growth 
of these remnants does not reproduce the idealized broken power-law radii 
of canonical SNe, chiefly because they do not exhibit self-similar behavior.  
Multiple reverse shocks reverberate throughout the remnant over its evolution, 
and it later violently collides with the dense H II region shell formed by 
its progenitor \citep[see also][]{byh03,ky05}.

If we adopt the conservative stance that a SN remnant halts star formation
in a halo as soon as it reaches it, and if the collapse time of the halo at 
the time of the explosion is known, Figure \ref{fig:25Ms} can be modified to 
include both radiative and kinetic feedback by the star. This is done simply 
by assuming that if a star can form in the evaporated halo in the absence of
an explosion, it will still form if central baryon collapse takes less time
than required for the shock to reach the halo.  The contraction of the cloud
from $n_c \sim$ 10 - 10$^3$ to 10$^8$ cm$^{-3}$ is leisurely, taking place
over 5 - 10 Myr \citep[e.g.][]{abn02,on07}. Thereafter, as three-body H$_2$
cooling takes over in the core of the halo, the formation of a star proceeds
very rapidly, in under a Myr.  We therefore take 8 Myr as the average time
for a partially evaporated halo to form a star.  Clearly, this estimate is 
approximate because it neglects the initial LW dissociation of the core, but 
it is reasonable given that our studies show that H$_2$ reforms there 100 - 
200 kyr after the death of the star.  

Arrival times for the shock at the center of the halo are complicated by 
the fact that they are a function of both the energy of the explosion and 
the mass of its host halo, as shown in Figure \ref{fig:rvst}. We tabulate 
minimum and maximum arrival times for the remnant at 150, 250 and 500 pc
in Table \ref{tbl:arr} using the 15 \Ms \ SN and 40 \Ms \ hypernova from
\citep{wet08a} as proxies for the 25 and 40 \Ms \ SNe in our study.  The
evolution of the 15 \Ms \ remnant is a reasonable approximation to that 
of the 25 \Ms \ remnant, given the uncertainty in explosion energy in 
primordial stars over this mass range.  The hypernova, whose explosive 
yield is ten times greater than that of a 40 \Ms \ core-collapse SN, 
gives a lower limit to the arrival times of the remnant to the halo. The 
dispersion in times with host halo mass is due to the extra baryons the 
remnant must sweep aside to reach the halo, which can vary by more than a 
factor of ten in mass. The host halos we have chosen, 6.9 $\times$ 10$^5$ 
\Ms \ and 2.1 $\times$ 10$^{6}$ M$_{\odot}$, bracket those in which the 
remnant actually escapes the halo since neither progenitor can ionize halos 
greater than 10$^7$ \Ms \ \citep{wet08a}.  Since the hypernova remnant, 
which is the fastest, travels at most 500 pc in 10 Myr, star formation at 
500 and 1000 pc in Figure \ref{fig:25Ms} is unchanged. 

\begin{figure*}
\epsscale{1.15}
\plottwo{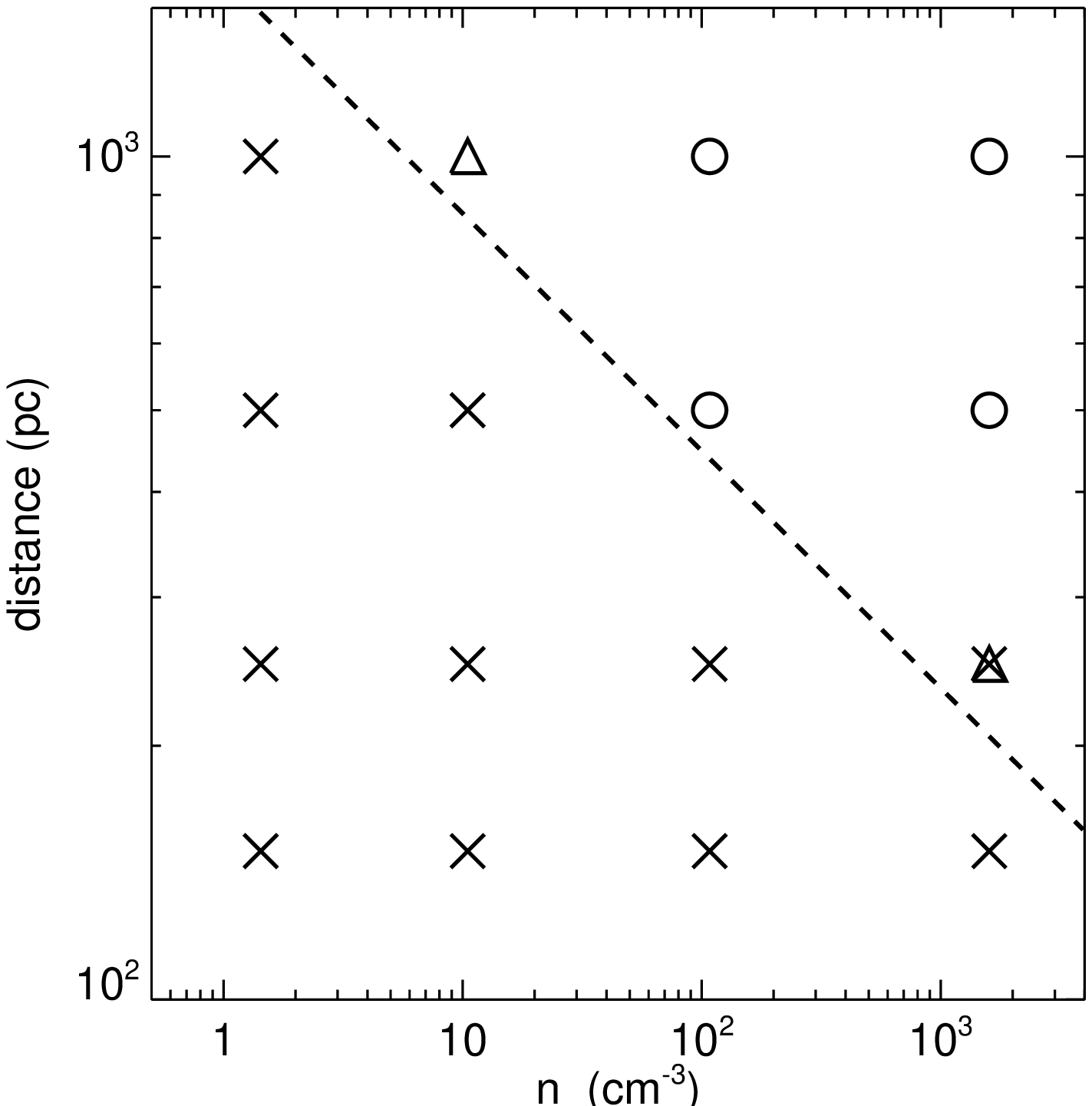}{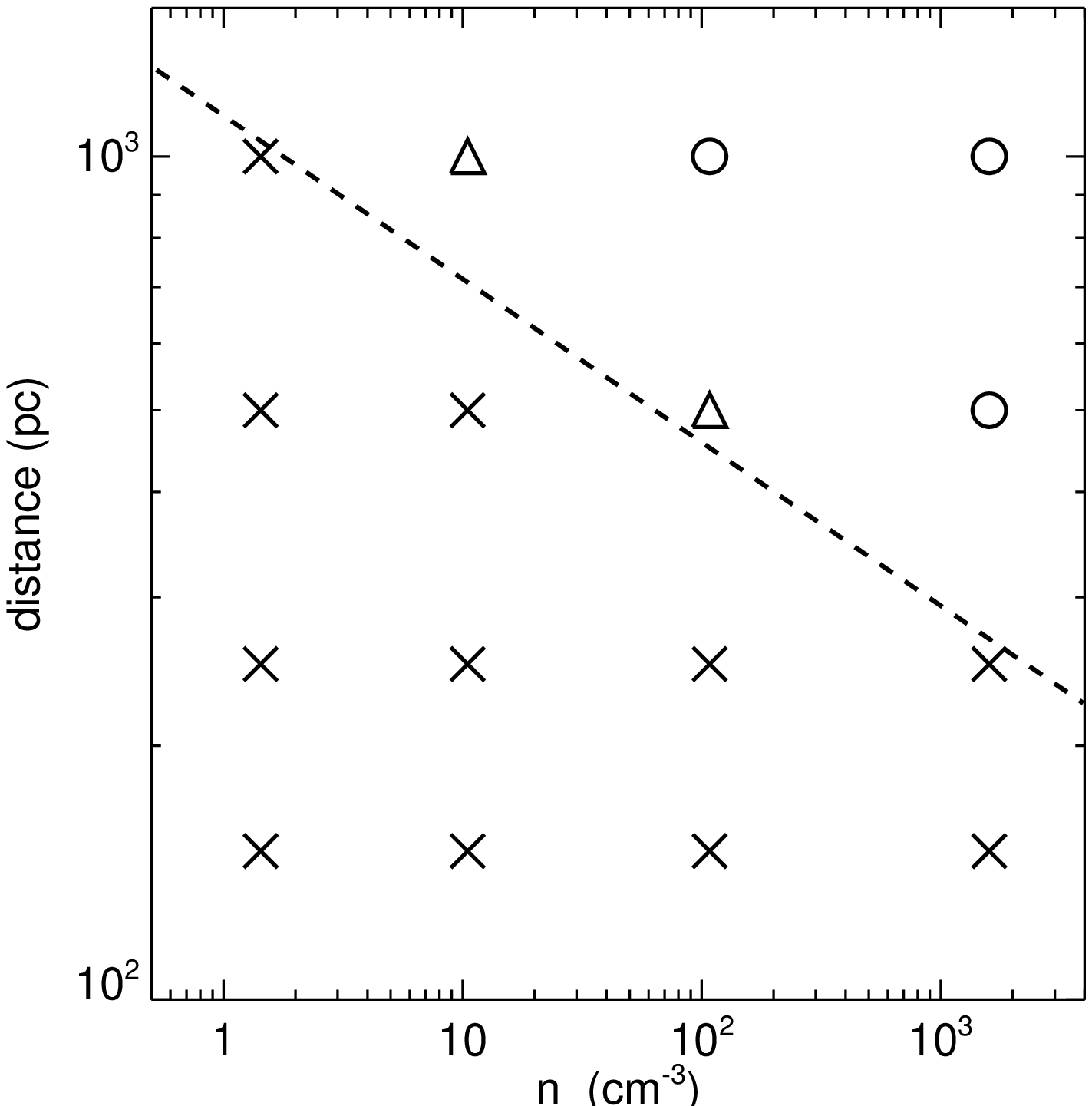}
\vspace{0.1in}
\caption{Radiative and kinetic feedback on star formation near a 25 \Ms \ 
star (left) and a 40 \Ms \ star (right). Completely evaporated halos with 
no star formation are labeled by crosses, while halos with delayed or
neutral star formation are marked by triangles and circles, respectively.  
The triangle overlaid on the cross in the 25 \Ms \ panel signifies that the 
halo can form a delayed star if the SN goes off in the 6.9 $\times$ 10$^5$ 
\Ms \ halo but not if it is in the 2.1 $\times$ 10$^{6}$ \Ms \ halo.  The 
dotted lines again define the boundary for star formation in the evaporated 
halos, above which it proceeds and below which it is quenched. 
\label{fig:new25Ms}} \vspace{0.1in}
\end{figure*}

We summarize both radiative and kinetic feedback by 25 and 40 \Ms \ 
stars in Figure \ref{fig:new25Ms}. Ram pressure stripping at most preempts 
delayed star formation in the densest halo ionized by the 25 \Ms \ star at 
150 and 250 pc; the less destructive of the 25 \Ms \ remnants only halts 
star formation at 150 pc.  On the other hand, both 40 \Ms \ remnants shut 
down delayed star formation at $n_c$ = 1596 cm$^{-3}$ at 150 and 250 pc. We 
find that kinetic feedback only modifies radiative feedback in halos at 400 
pc or less from low-mass stars.  The picture is more complicated with much
more energetic PISN, which can reach any halo within a typical cluster ($r 
\lesssim$ 1000 pc) in under 10 Myr \citep[Figure 10 of][]{wet08a}. Although 
the remnant overruns the halos sooner, radiative feedback can also speed up 
baryon collapse beforehand. Whether or not stars form in such cases remains 
unclear, and is beyond the scope of this study.

\begin{deluxetable}{rrrrr}
\tabletypesize{\scriptsize}
\tablecaption{SN Remnant Propagation Times \label{tbl:arr}}
\tablehead{
\colhead{SN} & \colhead{ halo ($M_{\odot}$)} & \colhead{ $t_{150 pc}$(yr)} & \colhead{ $t_{250 pc}$(yr)} & \colhead{ $t_{500 pc}$(yr)}}
\startdata
15 \Ms \  & 6.9E+05 & 1.3E+06 &     5.5E+06 & $>$ 1.0E+07 \\
15 \Ms \  & 2.1E+06 & 3.5E+06 & $>$ 1.0E+07 & $>$ 1.0E+07 \\
40 \Ms \  & 6.9E+05 & 3.1E+05 &     1.1E+06 &     8.0E+06 \\
40 \Ms \  & 2.1E+06 & 8.5E+05 &     2.4E+06 & $>$ 1.0E+07 \\
\enddata
\end{deluxetable}

We note that our estimates of kinetic feedback are conservative for 
several reasons. First, as stated earlier, we apply arrival times for a 
40 \Ms \ hypernova to both 25 and 40 \Ms \ core-collapse SNe, which have 
lower energies and velocities, and hence a shorter reach within the cluster. 
Second, we assume that the halos in our study are not connected by filaments 
of dark matter and gas, which is usually not the case.  If both the UV and 
the SN shock must propagate along a cosmological filament, with overdensities 
of 50 or more above the cosmic mean, to reach the core of a nearby halo they 
may have considerably less impact on subsequent star formation there. Finally, 
we also neglect the collision of the remnant with supersonic backflow from the 
evaporated halo, which will dampen its impact with the core of the halo. Also, 
40 - 50 \Ms \ primordial stars may die in more exotic explosions such as 
hypernovae or collapsars that are either asymmetric or beamed \citep[e.g.][]
{Tominaga2007,Tominaga2009}.  However, the asphericity of such events lowers 
the likelihood that any one halo in a cluster would be struck by ejecta, so 
for simplicity we do not consider them here.  

\subsection{Analytical Fits to Local Star Formation at High Redshift}

In our previous study we made the distinction between positive, negative, 
and neutral feedback within in a cluster of halos.  Negative feedback meant 
that star formation is either delayed or suppressed, positive feedback meant 
that a star forms more quickly in the halo than when there is no radiation, 
and neutral feedback meant that a star forms as fast as it would without UV. 
However, what is ultimately of importance to early structure formation and 
the rise of the first stellar populations is whether or not a star forms in 
the halo, not its exact timing, which varies by less than a few Myr in the 
scenarios we have investigated. Since the final result of neutral, positive, 
or no feedback is the formation of a star, the distinction between these 
outcomes is not relevant. Hence, the rules we formulate delineate the basic
threshold for star formation in a halo exposed to UV flux and SN flows from 
nearby primordial stars.

In Figures \ref{fig:60Ms} and \ref{fig:new25Ms} we plot lines that denote
the threshold radius $r_{th}$ from the star above which the halo can form 
a star if it is at a central gas density $n_c$ and below which it cannot.
The threshold has the simple form
\begin{equation}
log_{10} r_{th} \; = \; a_1 \; + \; b_1 (log_{10} n_c \; + \; c_1), 
\vspace{0.05in}  
\end{equation}
where $r_{th}$ is in pc and $n_c$ is in cm$^{-3}$.  We tabulate $a_1, 
b_1$ and $c_1$ in Table \ref{tbl:coeff} for 25, 40, 60 and 80 \Ms \ 
stars, including both radiative and kinetic feedback for 25 and 40 
\Ms \ stars.  As shown in Figure \ref{fig:120Ms}, radiative feedback
by 120 \Ms \ stars is even simpler: if the satellite halo has central
densities greater than 100 cm$^{-3}$ a new star will form anywhere in
the cluster.  If not, its formation is suppressed.
\begin{deluxetable}{rrrrr}
\tabletypesize{\scriptsize}
\tablecaption{Analytical Fits to Radiative and Kinetic Feedback\label{tbl:coeff}}
\tablehead{
\colhead{$M_{\star}$} & \colhead{$a_1$} & \colhead{$b_1$} & \colhead{$c_1$} }
\startdata
25 \Ms \  & 3.3010 & -0.33474 &  -0.14613   \\
40 \Ms \  & 3.1761 & -0.49480 &   0.30103   \\
60 \Ms \  & 3.3010 & -0.33011 &  9.6910E-02 \\
80 \Ms \  & 3.3010 & -0.33084 &  -0.69897   \\
\enddata
\end{deluxetable}
\subsection{Ionization Front Instabilities}

\begin{figure*}
\epsscale{1.15}
\plotone{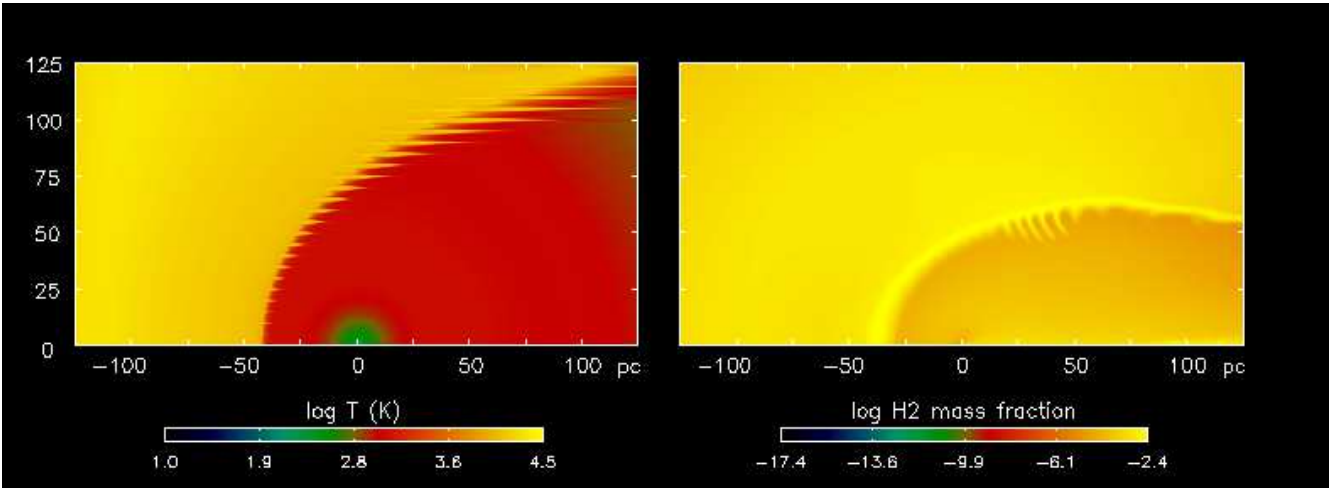}
\vspace{0.1in}
\caption{Temperatures at 600 kyr (left) and H$_2$ mass fractions at 10 Myr 
(right) in halo 073 ($n_c =$ 1596 cm$^{-3}$) 150 pc from a 25 \Ms \ star. 
\label{fig:H2instabil}} 
\vspace{0.1in}
\end{figure*}

Longer illumination times and lower fluxes promote the onset of dynamical
instabilities in the D-type ionization front as it engulfs the satellite 
halo, as we show for the 073 halo 150 pc away from a 25 \Ms \ primordial 
star in the left panel of Figure \ref{fig:H2instabil}. They arise because 
the I-front assumes a cometary shape and because the high energy tail of 
the spectrum forms H$_2$ between the front and the dense shell that 
radiatively cools the shell \citep[e.g.][]{rgs01,wn08b}.  Using rigorous 
perturbation analysis, \citet{rjr02} discovered that D-type fronts driven 
by photons that are oblique to the front are always unstable, and that the 
growth rates of the modes rise with the angle of incidence of the photons.  
This, together with cooling of the shocked shell by H$_2$, is the origin 
of the instabilities in the I-front enveloping the halo in Figure 
\ref{fig:H2instabil}. 

At early times, we find that the modes with the greatest amplitudes are 
indeed those furthest out along the arc of the I-front, where photons are
incident to the front at the greatest angles.  At intermediate and later 
times the perturbations grow nonlinearly and degenerate into turbulent 
fluid motion along the outer segments of the arc.  The amplitudes of the 
modes closest to the axis of the halo are small and instabilities never 
puncture its core. Much more prominent perturbations have been found in 
planar I-fronts approaching spherical molecular cloud cores in numerical 
models with efficient radiative cooling by molecules \citep{miz06}. These 
phenomena have been proposed for the origin of the ``Pillars of Creation'' 
in the Orion Nebula, but they are different from those in our simulations. 
They begin as Vishniac thin-shell overstabilities \citep{v83} caused by 
efficient molecular cooling in plane-parallel I-fronts, not curved ones, 
and later erupt into violent instabilities driven by ionizing radiation. 
In the \citet{miz06} models the unstable modes do propagate into the 
molecular cloud core.  This never occurs in our simulations because H$_2$ 
cooling is too inefficient to incite Vishniac modes.

We point out these features because they are prominent in many of our 
models, but they do not affect star formation in the halo because they 
never approach its inner regions.  Mostly, they just roil gas along the 
shock, breaking it up into clumps that can persist for up to 10 Myr. 
We find that they appear when the star is 150 or 250 pc from the halo and 
are most prominent with 25 and 40 \Ms \ stars.  Fewer instabilities appear
as stellar mass increases; they arise in only two of the 120 \Ms \ models 
and have lower amplitudes.  There are two reasons for this.  First, higher
mass stars have greater LW fluxes that lower H$_2$ cooling in the dense 
shell.  Second, larger ionizing UV fluxes result in shorter-lived cometary 
profiles in which unstable modes can develop. The arc is crushed downward 
into the shadow of the halo more quickly and the instabilities have less 
time to develop.

\section{Discussion and Conclusion}

We find that 25 - 120 M$_{\odot}$ primordial stars are relatively uniform
in their effect on new star formation within clusters of small halos at
high redshifts, before the rise of global LW backgrounds.  The evolution
in spectral profile from 25 - 120 \Ms \ has no impact on the formation of 
stars in nearby halos, which allows its removal from the parameter space
of local radiative feedback.  The empirical fits we have devised mark the 
threshold for star formation in satellite halos as a function of central 
baryon density, proximity to the star, and neighbor star mass.  Although 
the halo in our study is the just the least massive one found to form a 
star in previous AMR simulations, our results can be used as upper limits 
to feedback in more massive halos.  Our results imply that future surveys 
of local feedback with more massive halos can be accomplished with fewer 
stars, since outcomes for halo photoevaporation above and below the belt
in $n_c$ and radius in which there is variability is relatively uniform 
from 25 - 80 M$_{\odot}$.

Radiative and kinetic feedback between minihalos is key to many processes 
in early cosmological structure formation, such as primordial SNe event 
rates \citep{wa05}, especially those that account for cluster bias \citep{
mac06}, the rise of the first stellar populations, the assembly of primeval 
galaxies, and the evolution of metagalactic LW backgrounds.  Our analytical 
fits enable feedback estimates for a representative cut of Population III 
stars in analytical models of these early processes. They can also be used 
in numerical simulations, especially those performed in large cosmological 
boxes capable of resolving minihalo clustering but not of capturing halo 
photoevaporation.

Local radiative feedback at slightly lower redshifts is different due to 
the presence of LW backgrounds from primordial stars, which until recently
has been thought to be quite destructive to new star formation mediated by 
H$_2$ cooling in halos \citep{hrl97,har00,yahs03}.  However, recent, more
detailed simulations reveal that star formation in cosmological halos is 
postponed rather than prevented in LW backgrounds, even large ones that are 
consistent with a fully reionized universe \citep{wa07,on08}. \citet{wa07} 
found that a halo that formed a star at 5 $\times$ 10$^{5}$ \Ms \ in the 
absence of a photodissociative background still formed one by H$_2$ cooling 
50 Myr later after it grew by mergers and accretion to 5 $\times$ 10$^{6}$ 
\Ms \ in a uniform LW field of 1 $J_{21}$ ($=$ 10$^{-21}$ erg cm$^{-2}$ Hz$
^{-1}$ str$^{-1}$ s$^{-1}$), that of a fully-ionized universe.  Other halos 
in its vicinity also grew to larger masses, even though cooling and collapse 
of baryons were temporarily stalled in them.  Thus, at lower redshifts local 
UV feedback still begins when one star irradiates neighbor halos. In 
contrast to the first generation, LW backgrounds may continue to suppress 
star formation in photoevaporated halos after the death of the star by 
slowing the formation of H$_2$ in the relic H II region, the remnant 
I-front shock, and the halo core.

However, this effect may have been overestimated in previous analyses.
Consider the morphology and large densities and H$_2$ fractions in the 
relic H II region at 10 Myr in the right panel of Figure 
\ref{fig:H2instabil}.  Similar H$_2$ fractions would likely persist in 
the recombining H II region even in large LW backgrounds because the 
very high electron fractions there restore it so quickly via the H$^-$ 
channel.  This is especially true in the high densities of the I-front 
shock remnant, that can be seen to envelope the core in Figure 
\ref{fig:H2instabil}. Enough molecular hydrogen could be catalyzed in 
the envelope to expel the LW background from its interior, allowing 
H$_2$ to reform at the center of the halo, cool it, and form a star.  
Thus, photoevaporation may actually free satellite halos to form stars 
that were previously suppressed by the background.  This, together 
with our current study, suggests that star formation in cosmological 
halos at lower redshift was much more robust than is often supposed. 
Numerical models are now being developed to investigate the survival 
of molecular hydrogen, and therefore new star formation, in evaporated 
halos for a range LW backgrounds.

The effect of photoevaporation on the final mass of any star that does 
form in the halo is not yet well understood, but initial estimates by
\citet{suh09} suggest that it will be smaller than in undisturbed halos.  
They find that outflows and shock disruption in the core lower central 
accretion rates, and by extrapolating these rates from early stages of 
collapse out to Kelvin-Helmholtz contraction time scales, they conclude 
that the final star will be 25 - 50 M$_{\odot}$.  This mass scale is
similar to those on which Pop III.2 stars form due to HD cooling in 
relic H II regions \citep{yet07}.  HD is important because it can cool 
primordial H II regions down to the CMB temperature and lower the mass 
scales on which they fragment.  However, we do not include it in our 
models because it forms primarily in the relic ionized gas surrounding 
the halo core, not in the core itself.
 
Although our study is a significant extension to our earlier survey 
of local radiative feedback, additional feedback channels remain to 
be properly investigated. If the death of the star results in a black 
hole, accretion would expose nearby halos to its soft x-ray flux 
\citep{mba03,awa09}, creating significant free electron fractions in 
them due to secondary ionizations without strongly heating them. This 
process could enhance their H$_2$ mass fractions and promote their 
collapse into new stars. Likewise, the impact of SN ejecta with a halo 
that has been partially stripped by supersonic flows could deposit
metals into its interior and accelerate its cooling and collapse.  
These potential avenues of positive feedback on primordial star 
formation will be the focus of future simulations.

\acknowledgments

This work was carried out under the auspices of the National Nuclear Security Administration 
of the U.S. Department of Energy at Los Alamos National Laboratory under Contract No. 
DE-AC52-06NA25396.  The simulations were performed on the open cluster Coyote at Los Alamos 
National Laboratory.

\bibliographystyle{apj.bst}
\bibliography{ms.bib}

\end{document}